\documentclass[%
 reprint,
 amsmath,amssymb,
 aps,
]{revtex4-1}

\usepackage{graphicx}% Include figure files
\usepackage{dcolumn}% Align table columns on decimal point
\usepackage{bm}% bold math
\usepackage[T1]{fontenc}
\usepackage{multirow}
\usepackage{url}
\usepackage{xcolor} %% added by Kanish
\usepackage{epstopdf}
\begin{document}

%\preprint{APS/123-QED}
%\title{ Fundamental, technical and external factors induced positive and negative extreme events in stock market and its prediction}
%\title{Identification and prediction of extreme event in the stock market: Fundamental, technical and external factors induced positive and negative extreme events}
%\title{Identification and machine- learning based prediction of stock price movements during extreme events}
\title{Detection and Forecasting of Extreme event in Stock Price Triggered by Fundamental, Technical, and External Factors}
%\title{Identification of extreme events due to fundamental, technical and external factors and prediction of stock price movement during extreme event}
\author{Anish Rai, Salam Rabindrajit Luwang and Md Nurujjaman}
\email{md.nurujjaman@nitsikkim.ac.in}
\affiliation{Department of Physics, National Institute of Technology Sikkim, Sikkim, India-737139.}
\author{Chittaranjan Hens}
\affiliation{Center for Computational Natural Sciences and Bioinformatics, International Institute of Information Technology, Hyderabad, India-500032.}%
\author{Pratyay Kuila}
\affiliation{Department of Computer Science, National Institute of Technology Sikkim, Sikkim, India-737139.}%
\author{Kanish Debnath}
\affiliation{School of Business, RV University, Bengaluru, India-560059.}%

\date{\today}% It is always \today, today,

\begin{abstract}

The sporadic large fluctuations are seen in the stock market due to changes in fundamental parameters, technical setups, and external factors. These large fluctuations are termed as Extreme Events (EE). The EEs may be positive or negative depending on the impact of these factors. During such events, the stock price time series is found to be nonstationary.  Hence, the Hilbert-Huang transformation (HHT) is used to identify EEs based on their high instantaneous energy ($IE$) concentration. The analysis shows that the $IE$ concentration in the stock price is very high during both positive and negative EE with $IE>E_{\mu}+4\sigma,$ where $E_{\mu}$ and $\sigma$ are the mean energy and standard deviation of energy, respectively. Further, support vector regression is used to predict the stock price during an EE, with the close price being found to be the most useful input than the open-high-low-close (OHLC) inputs. The maximum prediction accuracy for one step using close price and  OHLC prices are 95.98\% and 95.64\% respectively. Whereas, for the two steps prediction, the accuracies are 94.09\% and 93.58\% respectively. The EEs found from the predicted time series shows similar statistical characteristics  that obtained from original data. The analysis emphasizes the importance of monitoring factors that lead to EEs for an effective entry or exit strategy as investors can gain or lose significant amounts of capital due to these events.

\end{abstract}

\pacs{Valid PACS appear here}% PACS, the Physics and Astronomy
\maketitle

%\tableofcontents

\section{Introduction}

The sudden emergence of an extraordinary event from an ordinary state, such as floods, earthquakes, stock market crashes, power blackouts, heart attacks, global war, or pandemics can be termed an Extreme Event (EE)~\cite{bunde2012science,zhang2009estimating,albeverio2006extreme,mahata2021characteristics,banerjeepredicting,mishra2018dragon, karnatak2014route,albeverio2006extreme,malik2020rare,chowdhury2022extreme}. EE impacts a system and its surrounding in a severe way, and there are various factors that lead to an EE in a system. The analysis of different factors-induced EE is of utmost importance as such events have a very strong impact~\cite{ray2020extreme,malik2020rare}. Though the study of EEs has been carried out in various disciplines, very limited work has been done in finance, especially in the stock market~\cite{mcphillips2018defining,mahata2021characteristics}. Studying the identification and characterization of EEs in the stock market is crucial for understanding their impact. 

Generally, an EE leads to a sudden crash in certain stocks and as a result, the investors face huge losses~\cite{mahata2021characteristics,rai2022statistical}. But, the same EE has a positive impact on certain stocks and hence leads to a sudden stock price upsurge ~\cite{pal2010foreign,nagaraj1996india}. These upsurges result in huge gains for investors. Thus, categorizing the EE in the stock market according to its effect on the stock price is crucial. No prior research has categorized an EE based on the impact, yet it is crucial to do so. We can categorize the EE in following way. The EE that leads to a sudden crash in the stock price can be termed a negative  EE and or that the EE that leads to upsurge can be termed as positive EE. The stock market crash of 1987, the dot-com crash of 2000, the financial crash of 2008 and the recent COVID-19 crash in the stock market can be considered negative EEs~\cite{ seyhun1990overreaction, morris2012value, acharya2009causes,mahata2021modeling,rai2022sentiment,rai2022statistical}. On the contrary, the stock market that experience rapid price surges due to fundamental and external factors, such as liberal policies, strong corporate results, favorable weather, and tax reductions, as seen in the BSE SENSEX in 1991 and S\&P Index in 1982 may be considered as positive EE~\cite{pal2010foreign,nagaraj1996india,monsoon}. 

Research on the impact of crashes and upsurges in stock prices on EE is limited. It is particularly important for investors to understand the company-specific EE associated with various factors when making entry or exit decisions in the market. Positive and negative EEs are influenced by fundamental, technical, and external factors, which are explained below.
\begin{itemize}
 \item{ \textbf{Fundamental Factors:}} Investors analyze a company's financial parameters, which are disclosed quarterly through financial results, as part of their evaluation of fundamental factors before making investments~\cite{gursida2017influence}. Significant changes in a company's net profit, sales, earnings, and growth, which indicate a significant shift in overall performance, have a strong impact on the movement of the stock price~\cite{emamgholipour2013effects,wang2013accounting}. Studies indicate that the stock price is highly responsive to major corporate decisions, changes in stake holdings, regulatory policies, macroeconomic conditions, brokerage firms' buy or sell recommendations, and other factors that affect a company's fundamentals~\cite{cutler1988moves,patell1982good}. Thus, these fundamental factors can result in either positive or negative EE.

\item{\textbf{Technical factors:}} Technical factors can significantly impact stock prices, as the formation of various chart patterns may cause sudden changes in prices~\cite{taylor1992use,wong2003rewarding}. The sudden change in stock price is caused by uninformed traders making synchronized buy or sell trades~\cite{barlevy2003rational,gennotte1990market}. Various chart patterns indicate distinct stock price trends, such as ascending or descending triangles, double-bottom, and support-resistance patterns. These technical chart patterns may cause sudden crash or upsurge in the stock price, resulting in both types of EEs ~\cite{leigh2002stock,bulkowski2021encyclopedia}. Occasional EE can significantly affect profitability, making it valuable for investors to study such events.
\item {\textbf{External Factors:}}External factors such as interest rate changes done  by the U.S. Federal Reserve, political events like elections, global events like wars and natural hazards like earthquakes can cause significant fluctuations in stock prices~\cite{petersen2010quantitative, nazir2014impact, mahmood2014impact, boungou2022impact, hudson2015war, schneider2006war, worthington2004measuring, bourdeau2017impact}. Therefore, analyzing the impact of these factors on stock price movements is necessary to identify an EE.
\end{itemize}

Our paper provides an analysis of characteristics and predicts stock prices during both positive and negative EEs. These events can be characterized by their magnitude or quantified impact~\cite{mcphillips2018defining}.  Events that exceed four to ten times the standard deviation of normal events are typically classified as EEs, as noted in previous studies~\cite{mishra2020routes,chaurasia2020advent,mahata2021characteristics}. The same analysis can be performed by estimating the Instantaneous energy ($IE$) spectrum, given the occurrence of sharp falls or rises in stock prices during negative and positive EEs. Due to the nonstationary and nonlinear characteristics of the stock price time-series, one can estimate the $IE$ utilizing the Hilbert-Huang transformation ($HHT$) method based on Empirical mode decomposition (EMD). As suggested by Mahata et al. (2021) ~\cite{mahata2021characteristics}, an EE can be identified when its $IE$ exceeds $E_{\mu}+4\sigma$, where $E_{\mu}$ and $\sigma$ represent the mean and standard deviation of energy, respectively.

The prediction of the stock price, especially during an EE, is the most important challenge in the stock market. Using support vector regression (SVR), a supervised machine learning technique, we have taken the initial step towards predicting stock price movements during positive and negative EEs in the stock market~\cite{kim2003financial,khaidem2016predicting,huang2009hybrid, kumbure2022machine}. While many studies cover the use of machine learning algorithms, particularly support vector regression (SVR), to predict stock prices during normal periods, no current research exists on applying these methods to predict stock prices during an EE.  Here, we have given one and two step ahead prediction of stock price during an EE. This prediction with optimum accuracy may help a trader for short term trading.

\begin{table}[!ht]
\caption{ The table represents the terminology used and their description.}
\label{tab:terminology}
\begin{center}
\begin{tabular}{ |c|l| } 
 \hline
 \textbf{Terminology} & \textbf{Description }\\ 
\hline
 SVR & Support Vector Regression \\ 
 \hline
 EMD & Empirical Mode Decomposition \\
\hline
 HHT & Hilbert-Huang transformation \\
\hline
 DNS & Degree of nonstationarity  \\
 \hline
EE& Extreme Event\\
 \hline
1-step ahead prediction& Prediction of $(t+1)^{th}$ step using \\
& $t^{th}$ step features \\
 \hline
2-step ahead prediction &Prediction of $(t+2)^{th}$ step using \\
& $t^{th}$ step features \\
 \hline
30-block& Prediction of 30 steps using 1-step ahead\\
& or 2-step ahead prediction\\
 \hline
5-block& Prediction of 5 steps using 1-step ahead \\
&or 2-step ahead prediction\\
 \hline
1-block& Prediction of 1 step using 1-step ahead \\
&or 2-step ahead prediction\\
 \hline
MAPE& Mean Absolute Percentage Error \\
 \hline
OHLC& Open-High-Low-Close price\\
 \hline

\label{tab:abb}
\end{tabular}
\end{center}
\end{table}

Rest of the paper is organized as follows. Table~\ref{tab:terminology} contains the terminology used in this paper.  Methods used for the analysis are discussed in Section~\ref{sec:MOA}. The data analysed is represented in Section~\ref{sec:DA}. The results are discussed in Section~\ref{sec:rd}. Section~\ref{sec:con} consist of conclusion of the work.

\section{Method of Analysis}
\label{sec:MOA}

In order to identify Extreme Events, we applied several techniques, which are elaborated below. First, we assessed the stationarity of the stock price time-series by conducting the Augmented Dickey-Fuller (ADF) and degree of nonstationarity (DNS) tests. Next, we employed the Hilbert-Huang transformation (HHT) method based on Empirical mode decomposition (EMD) to estimate the $IE$ used for identifying EEs. Finally, we applied Support Vector Regression (SVR) to forecast stock prices during positive and negative EEs. Further information on each of these techniques can be found below.

\subsection{Augmented Dickey-Fuller (ADF) Test}
\label{sec:ADF}

The ADF test is a commonly used unit root statistical test for checking the stationarity of a time-series~\cite{mushtaq2011augmented}.This test is an extension of the Dickey-Fuller (DF) test, and characterizes the time-series using an Auto Regressive (1) model, as in the original DF test. The Augmented Dickey-Fuller (ADF) test was proposed by Said and Dickey to address the issue of autocorrelation in the original DF test. They achieved this by adding additional lag terms of the dependent variables to the model~\cite{said1984testing,herranz2017unit}. The regression involved in the ADF test is as follows:

\begin{equation}
\Delta y_t=\beta +\gamma t + \alpha y_{t-1} + \sum_{j=1}^{n} \psi_j \Delta y_{t-j} + \epsilon_t
\label{eqn:1}
\end{equation}

Where, $\Delta$ is the difference operator, $\epsilon_t$ is the white noise error, $\beta$ is the contant, $\gamma$ is the trend of the regression. 

The ADF test is a statistical hypothesis test used to determine whether a time-series has a unit root, which would indicate nonstationarity. The null hypothesis for the test is the presence of a unit root in the time-series, while the alternative hypothesis is that the time-series is stationary and has no unit root.~\cite{mushtaq2011augmented,paparoditis2018asymptotic}. In particular,
\begin{align}
H_o : \alpha =1 \Rightarrow y_t \sim {I}(1)\\
H_1 : |\alpha|<1 \Rightarrow y_t \sim {I}(0)
\end{align}

The t-statistics used to test the null hypothesis are based on the least squares estimates of Equation~\ref{eqn:1}, which is given by:
\begin{equation}
t_{\alpha=1} = \frac{\hat{\alpha}-1}{SE(\alpha)}
\label{eqn:2}
\end{equation}
Where, $\hat{\alpha}$ is the least square estimate and ${SE(\alpha)}$ is the standard error estimate. 

The ADF test provides several outputs, including the t-statistic, p-value, and critical value at different levels of significance. In the case of a nonstationary time-series, the t-statistics and p-value will be greater than the critical value and level of significance, respectively~\cite{kim2017unit,mushtaq2011augmented}.

\subsection{Empirical Mode Decomposition (EMD)}
\label{sec:EMD}

The empirical mode decomposition (EMD) technique is a widely used method for decomposing nonlinear and nonstationary time-series data, as described in several studies~\cite{huang1998empirical,mahata2020identification,mahata2020time}. This technique decomposes the time-series into unique intrinsic mode functions (IMFs), each of which has a definite timescale. To be considered an IMF, a time-series must satisfy two criteria:

\begin{enumerate}
\item the number of zero crossings and the number of extrema (i.e., maxima and minima) must be equal or differ by at most one.  
\item the mean value of the envelope defined by the local maxima and the envelope defined by the local minima must be zero.
\end{enumerate}

A detailed procedure to decompose a signal into $IMF$s are given below:

\begin{enumerate}
\item The upper envelope ($UE_t$) and lower envelope ($LE_t$) of a data set ($D_t$) are drawn by connecting maxima and minima, respectively, with the help of spline fitting.
\item A new dataset ($ND_t$) is obtained by subtracting the old dataset ($D_t$) with the mean value of the envelope, where $mean=(UE_t+LE_t)/2$. $ND_t=D_t-mean$. 
\item Untill the $IMF$ criteria are satisfied, the above steps (1) \& (2) are repeated taking $ND_t$ as the new dataset. After the conditions of $IMF$ are satisfied, the dataset $ND_t$ is considered as the first $IMF$ i.e., $IMF1$. 
\item To calculate the $IMF2$ of the dataset, the same process is followed with the dataset $N_t$, where $N_t=D_t-IMF1$. The decomposition process continues until the final dataset is monotonic. This monotonic dataset is called the residue and represents the overall trend of the original dataset, $D_t$. The original dataset can be reconstructed as the sum of all the $IMFs$ and the residue, that is, $D_t=\displaystyle \sum_{j=1}^n IMFj+residue$, where $IMFj$ represents the $j^{th}$ $IMF$ and $n$ is the total number of $IMFs$.
\end{enumerate}

Hilbert transform is used to calculate the instantaneous frequency ($\omega$) of a $IMF$. Hilbert transform is defined as $\displaystyle A_t=\frac{C}{\pi}\int_{-\infty}^{\infty}\frac{IMF}{t-t'}dt$. Here, $C$ is the value of Cauchy principle. $\omega$ is defined as $\displaystyle \omega=\frac{d\theta_t}{dt}$, where $\displaystyle \theta_t=tan^{-1}\frac{A_t}{IMF}$~\cite{huang1998empirical}. 

Hilbert spectrum [$H(t,~\omega)$] is a time-frequency distribution of a time-series and, is defined as

\begin{equation}
H(t,\omega)=Re\sum_{i} M_i(t)~e^{i\int{\omega_i(t)dt}}
\label{eqn:hs}
\end{equation}

Here, $M_i(t)$ is the amplitude and $\omega_i(t)$ is the frequency of the $i^{th}$ IMF. We have considered all the $IMFs$ excluding the residue to estimate the $H(t,~\omega)$ as residue is a monotonic function.

Instantaneous energy $IE(t)$, which is useful to identify EE, can be estimated from $H(t,~\omega)$ and it is defined as

\begin{equation}
IE(t)=\int_{\omega} H^2(t,\omega)d\omega
\label{eqn:ie}
\end{equation}

In our study, we applied Equation~\eqref{eqn:hs} to estimate the $H(t,\omega)$ of a combination of $IMFs$. By identifying the maximum energy concentration in the $H(t,\omega)$ spectrum, we can readily determine the occurrence time of an Extreme Event (EE).

However in order to quantify an EE we need to estimate the $IE$ of the combined $IMF$ from Eqn.~\eqref{eqn:ie}. We have used the normalized $IE_N(t)$ for the analysis, where $IE_N(t)=IE(t)/max[IE(t)]$. Events whose $IE_N$ are greater than the threshold value will be identified as an EE. The threshold energy is calculated as $E_{th} = E_\mu + B\sigma$, where $B$ is a constant, $E_{\mu}$ is the average energy and $\sigma$ is standard deviation of the energy. We have chosen B=4 for our analysis.

Degree of nonstationarity $[DNS(\omega)]$ test is applied to check the stationarity of a time-series. DNS is defined as

\begin{equation}
DNS(\omega)=\frac{1}{T}\int_{0}^{T}[1-\frac{H(t,\omega)}{h(\omega)}]dt
\label{eqn:dns}
\end{equation}

Here, [0, $T$] is the time window and $h(\omega)$ is the marginal spectrum and it is defined as, $h(\omega)=\frac{1}{T}\int_{0}^{T}H(t,\omega) dt$. In the case of stationary time-series, the $DNS(\omega)$ plot is a horizontal line and for nonstationary time-series data, the $DNS(\omega)$ plot is not a horizontal line~\cite{huang1998empirical}.

\subsection{Prediction Methods}
\label{sec:PM}
To predict the stock price during positive and negative EEs in the stock market, we employed the Support Vector Regression (SVR) machine learning technique. The following section outlines the methodology of this technique.

\subsubsection{Support Vector Regression}
\label{sec:SVR}

SVR is a supervised machine learning algorithm used for nonlinear regression~\cite{karimipour2019novel}. The methodology of SVR is similar to that of Support Vector Machines (SVMs). To train an SVR model, a set of input variables $(x_t,y_t)$ is required, which is also known as the training set. In this set, $x_i \in R^d, i=1,2,...,m$ represents the input variables, and $y_i \in R$ represents the corresponding targets. Using these input variables, a kernel-based regression model is trained to find the regression hyperplane."

SVR builds a regression function as:
$$f(x,w) = w^T\phi(x)+b$$

A margin of tolerance $\epsilon$ is set which is a threshold for error consideration.
\begin{equation}
\label{eq:1}
|y-f(x,w)|_\epsilon =\begin{cases}
																		0,~~~~~~~~~~~\text{if $|y-f(x,w)|\le \epsilon$}\\
\text{$|y-f(x,w)| - \epsilon$},~otherwise.  
\end{cases}  
\end{equation}

$||w||^2$ and sum of $\epsilon$-insensitivity are simultaneously minimized to estimate $f(x,w)$ as shown in Eq.~\eqref{eq:2}. C is a constant that controls the trade-off between an approximation error and ||w||.

\begin{equation}
\label{eq:2}
R = \frac{1}{2}||w||^2 + C(\sum_{i=1}^{m}|y-f(x,w)|_\epsilon)
\end{equation}

The minimization of risk shown in Eq.~\eqref{eq:2} is same as minimizing the risk in Eq.~\eqref{eq:3} under the conditions of Eq.~\eqref{eq:4}- \eqref{eq:5}. $\zeta_i$ and $\zeta_i^*$ are variables, one for exceeding and other for being below the target variable by more than $\epsilon$, respectively. $\zeta_i$ and $\zeta_i^* \geq 0$ and i=1,2,3....,m.

\begin{equation}
\label{eq:3}
 R = \frac{1}{2}||w||^2 + C(\sum_{i=1}^{m}(\zeta + \zeta_i^*)) 
\end{equation}

\begin{equation}
\label{eq:4}
 (w^T\phi(x)+b) - y_i \le ~\epsilon + \zeta_i 
\end{equation}

\begin{equation}
\label{eq:5}
 y_i - (w^T\phi(x)+b) \le ~\epsilon + \zeta_i^*
\end{equation}

A radial basis kernel function is utilized to map the input vectors $x_i \in R^d$ into a higher dimensional feature space. The function is defined as $K(x_i, x_j) = \exp(-\gamma||x_i-x_j||^2)$, where $\gamma$ is a parameter and $d$ is the degree of the polynomial function. The values of $\gamma$ and the regularization constant $C$ are selected using GridsearchCV. The analysis is conducted using a total of 600 data points, with 570 data points (95\%) being used for training and the remaining 30 data points (5\%) for testing the model. For first case, the input for the model is only the previous close price, and the model is used to predict the 1-step ahead and 2-step ahead close price for 30 blocks, 5 blocks, and 1 block. For 30 and 5 blocks, the model is used to predict the stock price in and around the occurrence of EE, while for 1 block, the model is used to predict the stock price precisely at the EE. For second case, the previous open-high-low-close (OHLC) prices are also utilized to predict the 1-step ahead and 2-step ahead close price for 30 blocks, 5 blocks, and 1 block.

\section{Data Analyzed}
\label{sec:DA}

We conducted an analysis of stock prices from companies based in various countries, including the USA, Germany, Russia, India, China, and the UK. Table~\ref{tab:1} shows the list of companies that we analyzed in this paper. We selected the length of the data based on the occurrence and duration of the event, as specified in table~\ref{tab:2}. We obtained all the data from the Yahoo Finance website~\cite{Yahoo}. The following sections present the results of our analysis.

\section{Results and discussion}
\label{sec:rd}

\begin{table}[!ht]
\tiny
    \centering
    \begin{tabular}{|l|r|r|r|}
    \hline
      \textbf{Company Name} &
      \textbf{t-statistic} &
      \textbf{critical Value} & 
      \textbf{p-value} \\
      \hline\hline
         Avast plc. & 0.8703 & -2.5756 & 0.8963 \\
       \hline
         NINE & -0.7878 & -2.5754 & 0.3648 \\
       \hline
         Jocil Ltd. & 0.4367 & -2.5839 & 0.8065  \\
       \hline
        JUBLINGREA & -0.1589 & -2.5961 & 0.5904 \\
       \hline
         GODREJCP & 0.1447 & -2.5700 & 0.7046 \\
         \hline
         PNBHOUSING & -0.9479 & -2.5804 & 0.3034 \\
         \hline
         GameStop Corp. & -1.1703 & -2.5790 & 0.2221 \\
         \hline
         IRCTC Ltd. & 0.3702 & -2.5753 & 0.7864 \\
         \hline
         PHIA& -1.1233 & -2.5753 & 0.2395 \\
         \hline
         Shopify Inc. & 0.9929 & -2.5753 & 0.9152 \\
         \hline
         BioNTech SE & 0.6323 & -2.5874 & 0.8512 \\
         \hline
         9939.HK & -0.6831 & -2.5745 & 0.4011\\
         \hline
         ZEEL & 1.2775 & -2.5788 & 0.9487 \\
         \hline
         SELAN & 1.0176 & -2.5754 & 0.9187 \\
         \hline
          Netflix Inc. & -1.1157 & -2.5755 & 0.2424 \\
         \hline
         NK Rosneft' PAO & -0.3620 & -2.5756 & 0.5186 \\
         \hline
         Chevron Corp. & 1.9830 & -2.5890 & 0.9953 \\
         \hline
         HDFCLIFE & -0.8743 & -2.5753 & 0.3307 \\
         \hline
         Ishanch & 0.0894 & -2.5700 & 0.6843 \\
         \hline
         HITECHCORP & 1.9396 & -2.5774 & 0.9873 \\
         \hline
        TEXMOPIPES & 0.4454 & -2.5796 & 0.8090 \\
         \hline
         HAPPSTMNDS& 1.8528 & -2.5811 & 0.9845 \\
         \hline
				GLENMARK & 0.6507 & -2.5753 & 0.8557 \\
       \hline
			  Pinduoduo Inc.& -0.8401 & -2.5920 &0.3417  \\
       \hline
			ALKYLAMINE& 2.2430 & -2.5758 &0.9940  \\
       \hline
			Coal India Ltd.& 0.0963 & -2.5854 &0.6851  \\
       \hline
			ADANIPORTS& -0.3625 &-2.5700  &0.5187  \\
       \hline
			OXY&2.3747  &-2.5890  &0.9953  \\
       \hline
			Bajaj Finance Ltd.&1.2868  &-2.5898  &0.9490  \\
       \hline
			NVIDIA Corp.&	1.4612&-2.5754	&0.9643	\\
			\hline
		Macy's Inc.& 0.3787	&-2.5928	&0.7875	\\
			\hline
			AT\&T Inc.&0.0211	&-2.5810	&0.6581	\\
			\hline
			Boeing Co.&	0.7192&-2.5803	&0.8691	\\
			\hline
			
				\end{tabular}
    \caption{The table represents the t-statistic, critical value and p-value, respectively, obtained from the Augmented Dickey-Fuller test at 1\% significance level for all the companies. Abbreviation of a few companies: NINE=Nine Energy Service Inc., JUBLINGREA=Jubilant Ingrevia Ltd., GODREJCP=Godrej Consumer Products Ltd., PNBHOUSING=PNB Housing Finance Ltd., IRCTC Ltd.=Indian Railway Ctrng nd Trsm Corp Ltd., PHIA=Koninklijke Philips NV, 9939.HK=Kintor Pharmaceutical Ltd., ZEEL=Zee Entertainment Enterprises Ltd., SELAN=Selan Exploration Technology Ltd., HDFCLIFE=HDFC Life Insurance Company Ltd., Ishanch=Ishan Dyes And Chemicals Ltd., HITECHCORP=Hitech Corporation Ltd., TEXMOPIPES= Texmo Pipes and products Ltd., HAPPSTMNDS=Happiest Minds Technologies Ltd., GLENMARK=Glenmark Pharmaceuticals Ltd., ALKYLAMINE=Alkyl Amines Chemicals Ltd., ADANIPORTS=Adani Ports and Special Economic Zone Ltd., OXY=Occidental Petroleum Corp.}
    \label{tab:1}
\end{table}		
		
\subsection{Nonstationarity Tests}
\label{sec:ns}

We have carried out the stationarity test for all the stock price time-series using Augmented Dickey-Fuller (ADF) test as described in Sec.~\ref{sec:ADF}. Table~\ref{tab:1} presents the test results, indicating that the t-statistic values for all time-series are greater than their respective critical values. Additionally, the p-values for all time-series are also greater than the significance level of 0.01, as noted in previous studies~\cite{kim2017unit}. Therefore, the results suggest that the stock price time-series are nonstationary, meaning that the null hypothesis cannot be rejected and that the time-series have a unit root, as discussed in Sec.~\ref{sec:ADF}. As the stock price time-series is nonstationary, we use the $EMD$ based $HHT$ technique as discussed in Sec.~\ref{sec:EMD} for subsequent analysis. 

\begin{figure}[hbt!]
\includegraphics[angle=0, width=8cm, height=7cm]{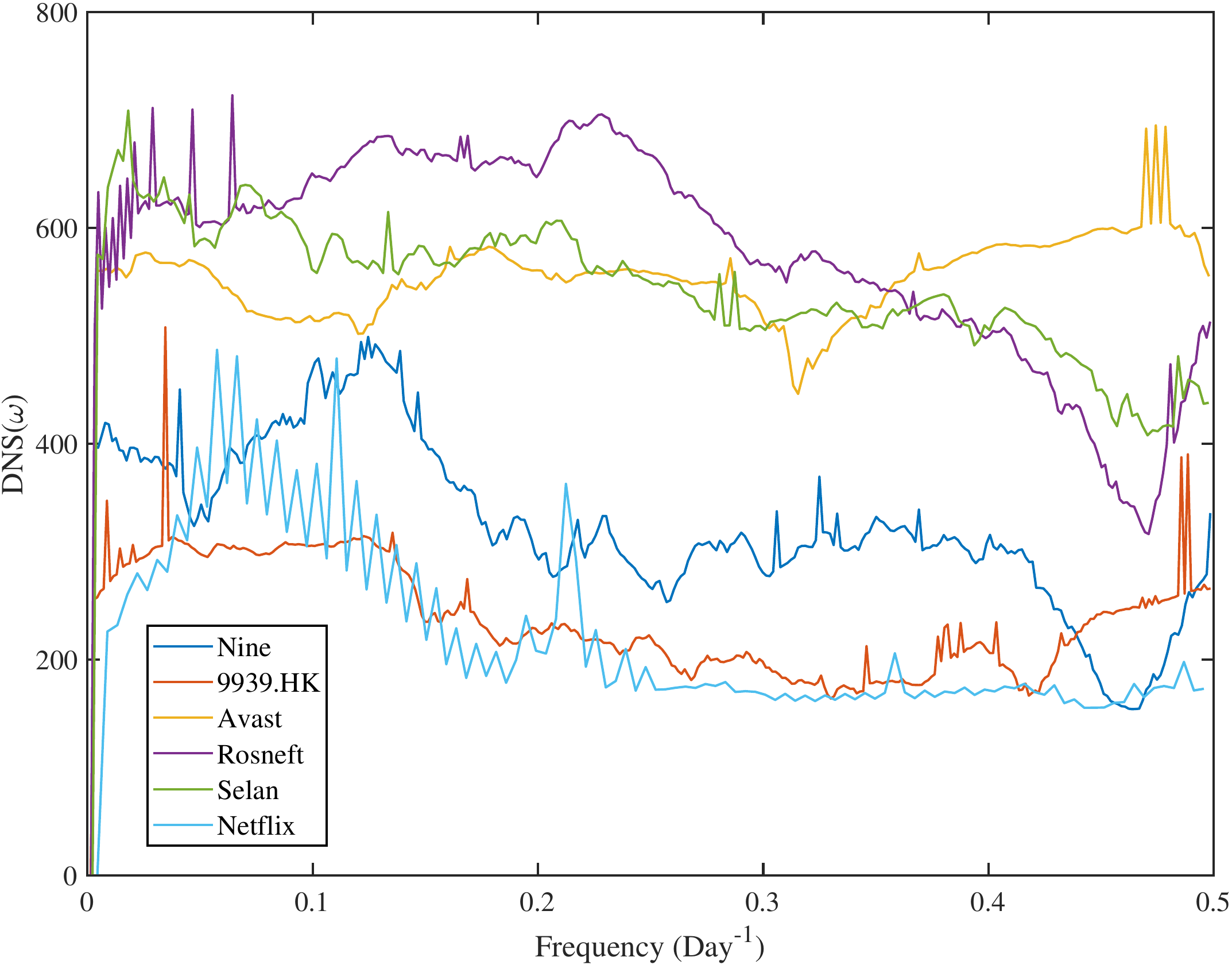}
\caption{\label{fig:DNS_plot_daily}The figure represents the DNS plot of Nine Energy Service Inc., Kintor Pharmaceutical Ltd.(9939.HK), Avast Plc., NK Rosneft' PAO, Selan Exploration Technology Ltd. and Netflix Inc., respectively.}
\end{figure}

In order to validate the result of the ADF test, we further use degree of nonstationarity (DNS) test as described in Sec.~\ref{sec:EMD}. Fig.~\ref{fig:DNS_plot_daily} represents the $DNS(\omega)$ plots of Nine Energy Service Inc., Kintor Pharmaceutical Ltd.(9939.HK), Avast Plc., NK Rosneft' PAO, Selan Exploration Technology Ltd., Netflix Inc., respectively. The non-horizontal DNS($\omega$) plot is an indication of nonstationarity in the time-series. In this study, the DNS test was conducted on all the stock price time-series, and the results confirm that the time-series are indeed nonstationary. Therefore, the use of the EMD technique for analyzing EE in the stock price is appropriate.

\subsection{EE induced by Different Factors}
\label{sec:df}
Detailed results on the emergence of EE in the stock market due to fundamental, technical, and external factors are presented in the following sections.

\subsubsection{Fundamental factors}

\begin{figure}[hbt!]
\includegraphics[angle=0, width=9cm, height=7.5cm]{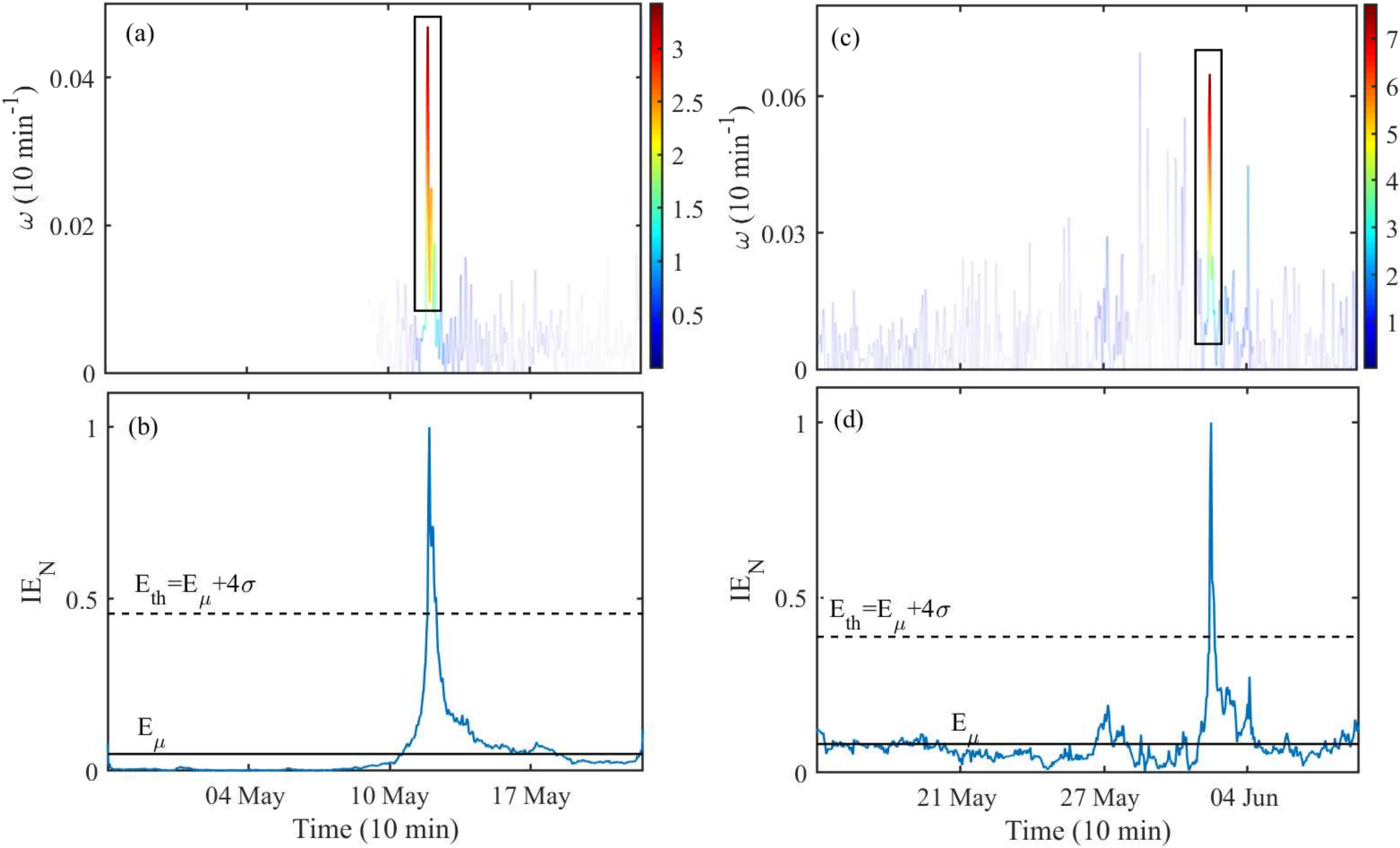}
\caption{\label{fig:Godrej_Ishanch_result}Plot (a) and (b) represent the $H(t,\omega)$ and $IE_N(t)$ plot of Godrej Consumer Products Ltd. and plot (c) and (d) represent the $H(t,\omega)$ and $IE_N(t)$ plot of Ishan Dyes and Chemicals Ltd., respectively.}
\end{figure}

The performance of a company is a crucial fundamental factor that investors rely on to make informed investment decisions. On May 12, 2021, the stock price of Godrej Consumer Products Ltd experienced a sudden price surge of approximately 28\% within minutes. This unexpected positive price movement was attributed to the release of positive financial results by the company. Fig.~\ref{fig:Godrej_Ishanch_result}(a) represents the Hilbert spectrum [$H(t,\omega)$] of the stock price of all the intrinsic mode functions ($IMFs$).  We have taken 10 min. data from 28 Apr. 2021 to 20 May 2021 for the analysis. The stock price surge is clearly captured by $H(t,\omega)$ as shown by the red patch in Fig.~\ref{fig:Godrej_Ishanch_result}(a). The red patch represents the maximum energy concentration. Further, normalized instantaneous energy [$IE_N(t)$] of the stock price time-series is shown in fig.~\ref{fig:Godrej_Ishanch_result}(b). It shows a clear spike in $IE_N(t)$ during the event. The $IE_N(t)$, the average energy ($E_{\mu}$) and threshold energy ($E_{th}$) are represented by blue solid, black solid and dotted lines, respectively, in Fig.~\ref{fig:Godrej_Ishanch_result}(b). As discussed in Sec.~\ref{sec:EMD}, the events with $IE_N$ greater than $E_{th}$ are considered as an EE, where $E_{th}=E_{\mu}+4\sigma$. $\sigma$ is the standard deviation of the energy. As the $IE_N$ during the event exceeded the $E_{th}$ threshold, the price surge of Godrej Consumer Products Ltd. due to its excellent financial results is considered a positive EE. We have identified similar positive EEs for several other stocks that have unexpectedly performed well which is shown in Table~\ref{tab:2}.

Conversely, the results of Ishan Dye and Chemicals Ltd. were announced on May 31, 2021, and were below market expectations~\cite{ishan}. Consequently, the stock price experienced an almost instantaneous crash of around 20\%. Fig.~\ref{fig:Godrej_Ishanch_result}(c) shows the $H(t,\omega)$ of the stock price. For this analysis, we have taken 10 min data from 17 May 2021 to 04 June 2021. The crash in the stock price is captured by the red patch in the $H(t,\omega)$. The $IE_N(t)$ of the above stock price time-series is also obtained from $H(t,\omega)$ and is shown in Fig.~\ref{fig:Godrej_Ishanch_result}(d). A clear spike is visible in the $IE_N(t)$ during the event. As discussed above, $IE>E_{\mu}+4\sigma$ for Ishan dyes and chemicals Ltd. due to unexpected bad result. Hence, the price crash can be considered as negative EE. Similar negative EE due to unexpectedly bad result are shown in Table~\ref{tab:2}. 
 
Positive and negative EEs due to unexpected good or bad results occur almost instantaneously, making it difficult for investors to enter or exit during these events. However, monitoring a company's financial position can help anticipate positive EEs and make early entries, while unexpected negative EEs may present entry opportunities at lower levels. Investors should keep some cash available for such opportunities.

\begin{figure}[hbt!]
\includegraphics[angle=0, width=9cm, height=7.5cm]{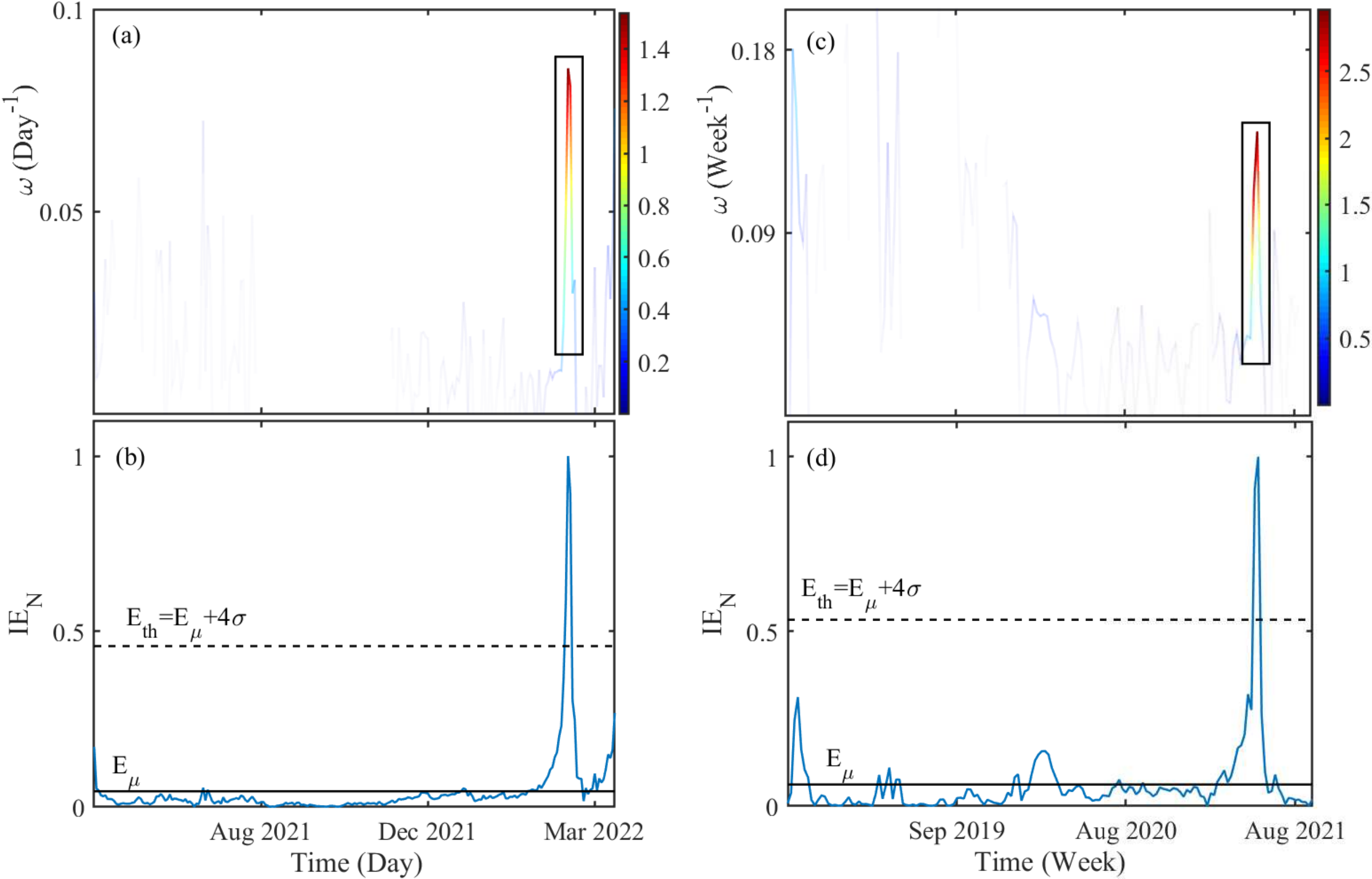}
\caption{\label{fig:NineD_PNBW}Plot (a) and (b) represent the $H(t,\omega)$ and $IE_N(t)$ plot of Nine Energy Service Inc. and plot (c) and (d) represent the $H(t,\omega)$ and $IE_N(t)$ plot of PNB Housing Finance Ltd., respectively.}
\end{figure}

Nine Energy Service Inc. has stated that their business activity has been steadily improving since the beginning of the year and they anticipate even better results in the future. As a result of their sustained growth and optimistic outlook, the stock price of Nine Energy Service experienced a significant increase. Fig.~\ref{fig:NineD_PNBW}(a) shows the $H(t,\omega)$ of the stock price of all the $IMFs$. The maximum energy concentration, shown by a red patch, is clearly visible during the stock price upsurge due to the positive outlook. Fig.~\ref{fig:NineD_PNBW}(b) shows the $IE_N(t)$ plot and a sudden spike is seen in $IE_N$ during the time of event. Since the spike exceeds the $E_{th}$, Nine Energy Service Inc.'s increase in stock price, attributed to positive growth and outlook, is regarded as a positive EE. The analysis utilizes stock price data with daily intervals ranging from May 2021 to April 2022.

In another example, On May 31, 2021, PNB Housing Finance Ltd.'s stock price surged rapidly following news of a private equity firm, Carlyle Group, acquiring over 50\% stake in the company. The graph depicted in Fig.~\ref{fig:NineD_PNBW}(c) shows the $H(t,\omega)$ for the stock price, indicating a distinct peak in energy concentration during a sudden upsurge in stock price. It further illustrates the $IE_N(t)$ obtained from the $H(t,\omega)$, which reveals a notable surge in $IE_N$ during this event, surpassing the threshold energy $E_{th}$. Therefore, the surge in PNB Housing Finance Ltd.'s stock price resulting from the acquisition of a stake can be regarded as a positive EE.

It is crucial to identify the underlying factors that cause an EE as it enables investors to adjust their entry or exit positions when faced with similar circumstances in the future. Based on the above results, it can be concluded that regular monitoring of the company's fundamentals and any new updates is essential in anticipating EEs. Through continuous monitoring, investors can make informed decisions about the appropriate entry or exit positions during EEs. To take advantage of EEs, investors should maintain some cash reserves.

\subsubsection{Technical factors}
\begin{figure}[hbt!]
\includegraphics[angle=0, width=9cm, height=10cm]{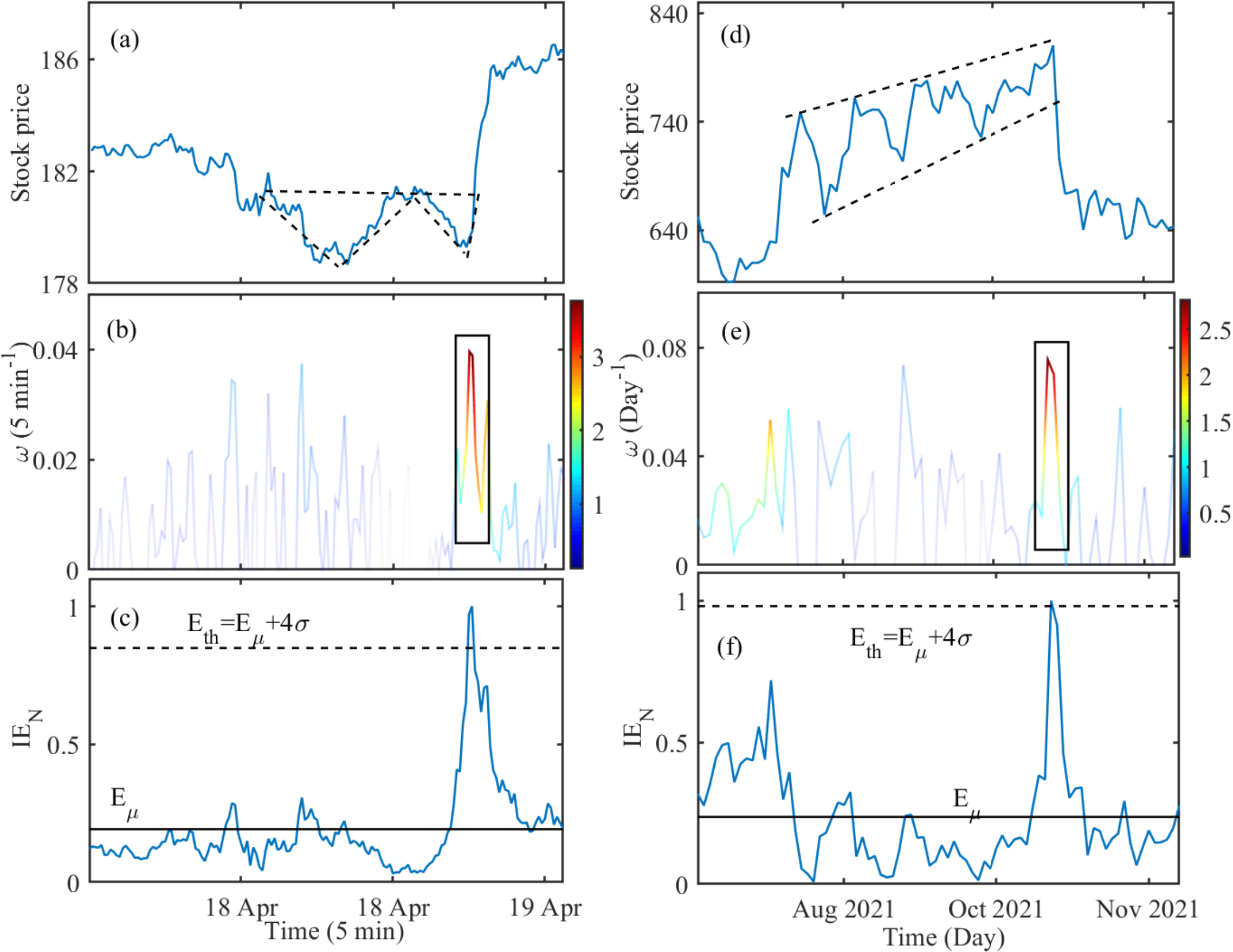}
\caption{\label{fig:boeing_jublinat_pattern}Plot (a), (b) and (c) represent the 5 min closing price, $H(t,\omega)$ and $IE_N(t)$ plot of  of Boeing Co. Plot (e), (f) and (g) represent the daily closing price, $H(t,\omega)$ and $IE_N(t)$ plot of Jubilant Ingrevia Ltd., respectively.}
\end{figure}
 
Technical analysis is a commonly used method in the stock market to generate profitable trades. Fig.~\ref{fig:boeing_jublinat_pattern} (a) displays the closing price of Boeing Co. between April 14, 2022, and April 19, 2022, at 5-minute intervals. During this period, a double-bottom (DB) technical chart pattern was formed, as indicated by the dashed lines in Fig.~\ref{fig:boeing_jublinat_pattern} (a). Following the formation of the DB pattern, a positive breakout in the stock price was observed. Fig.~\ref{fig:boeing_jublinat_pattern} shows a distinct concentration of maximum energy in $H(t,\omega)$ during the breakout, highlighted by the red patch. Fig.~\ref{fig:boeing_jublinat_pattern}  illustrates $IE_N(t)$ obtained from $H(t,\omega)$, revealing a significant spike in $IE_N(t)$ during the positive breakout, exceeding the threshold value $E_{th}$. Consequently, the breakout resulting from the DB chart pattern can be regarded as a positive EE. Similar positive breakouts in some other companies are presented in Table~\ref{tab:2}.

We have also observed the negative breakouts due to technical chart patterns. Fig.~\ref{fig:boeing_jublinat_pattern} (d) shows a daily closing price of Jubilant Ingrevia Ltd. from Jul. 2021 to Nov. 2021. A channel chart pattern was formed during this period  and it is shown by  dashed lines in Fig.~\ref{fig:boeing_jublinat_pattern} (d). A negative breakout is seen after the stock price broke the lower support line. Fig.~\ref{fig:boeing_jublinat_pattern} (e) shows that during a negative breakout the energy concentration is maximum. The breakout energy concentration is represented by red color. $IE_N(t)$ plot in Fig.~\ref{fig:boeing_jublinat_pattern} (f) also shows a sudden spike during the breakout. As the magnitude of the sudden spike is greater than the $E_{th}$, the negative breakout due to channel formation may be consdiered as a negative EE.

The results indicate that positive and negative EE can occur from various breakout patterns. Traders with technical knowledge can benefit from these EEs by entering during a bullish chart setup and exiting during a bearish setup. It's recommended to keep unallocated funds for entering before a breakout and follow a strict target and stop-loss to manage profits and minimize losses. Table~\ref{tab:2} shows EEs from different chart patterns. Finally, both types of EEs due to external factors are presented in the next subsection.

\subsubsection{External factors}

\begin{figure}[hbt!]
\includegraphics[angle=0, width=9cm, height=7.5cm]{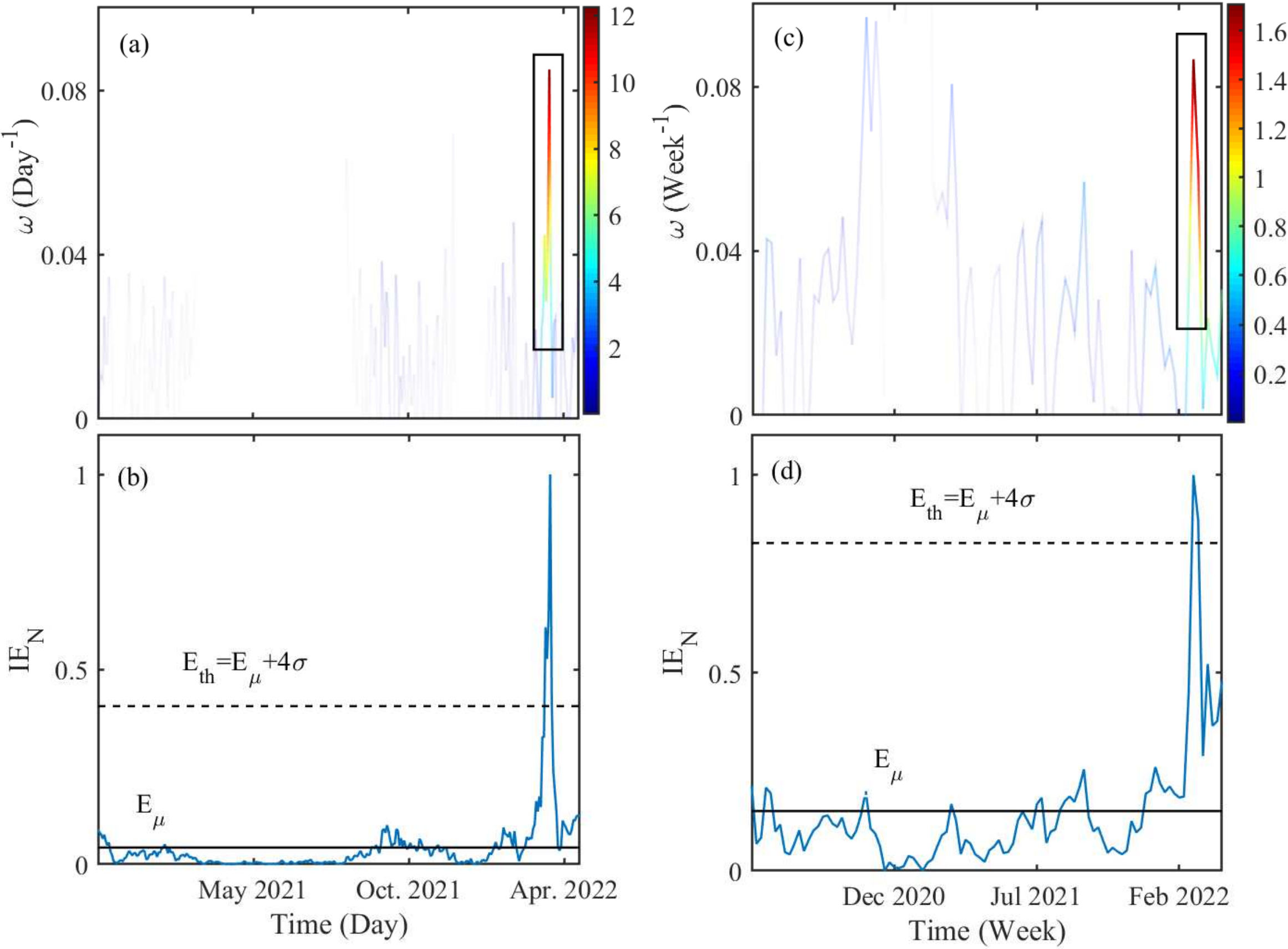}
\caption{\label{fig:Rosneft_chevron}Plot (a) and (b) represent the $H(t,\omega)$ and $IE_N(t)$ plot of NK Rosneft' PAO and plot (c) and (d) represent the $H(t,\omega)$ and $IE_N(t)$ plot of Chevron Corp., respectively.}
\end{figure}

On 24 Feb. 2022, Russia invaded Ukraine,  and due to this invasion, the Russian stock market crashed severely on the same day~\cite{Rusukraine,war}. For example, the stock price of NK Rosneft' PAO crashed nearly 60\% on the day of invasion. Fig.~\ref{fig:Rosneft_chevron} (a) represents the $H(t,\omega)$ of the stock price of all the $IMFs$ and it captures the crash due to Russia-Ukraine war by a red patch (maximum energy concentration). Fig.~\ref{fig:Rosneft_chevron} (b) represents the $IE_N(t)$ obtained from the $H(t,\omega)$. Fig.~\ref{fig:Rosneft_chevron} (b) shows a significant spike in $IE_N$ during the war which is greater than the $E_{th}$. Hence, the crash in the stock price of NK Rosneft' PAO due to Russia-Ukraine war is termed as a negative EE. 

In contrast, only a handful of companies profited from the Russia-Ukraine conflict. Following the invasion, some nations ceased importing oil from Russia, causing oil prices to skyrocket. As Russia is the second-largest exporter of oil, this resulted in a surge in global oil demand. As a result, the stock prices of oil corporations such as Chevron Corp. and Occidental Petroleum Corp. experienced a dramatic increase. For the analysis, the stock price data of Chevron Corp. was collected on a weekly basis, spanning from May 2020 to April 2022. The stock price upsurge is clearly captured by $H(t,\omega)$ as shown by the red patch (maximum energy concentration) in Fig.~\ref{fig:Rosneft_chevron} (c). Fig.~\ref{fig:Rosneft_chevron} (d) shows the $IE_N(t)$ of the stock price time-series that is obtained from the $H(t,\omega)$. It clearly shows an abrupt spike in the $IE_N$ during the war. This spike in $IE_N$ is greater than the $E_{th}$ and hence, the stock price upsurge in Chevron Corp. due to Russia-Ukraine war is considered as a positive EE.
 
The above results show that the external factor like war leads to both positive and negative EEs. In case of Russia-Ukraine war, stocks like NK Rosneft' PAO crashed heavily but stocks like Chevron Corp. upsurged rapidly. Such EEs provide opportunity to the investors to restructure their portfolio and invest in stocks with better future outlook. Therefore, investors should stay updated with the external factors like geopolitical news so that they can use such factors in their favor. EEs due to other external factors are shown in Table~\ref{tab:2}.

\begin{table*}[!ht]
\caption{ Col-2 represents the companies analysed, Col-3 represents the time-scale of the stock price data of each company, Col-4 shows the event that had occured, Col-5 shows the impact of the event on the stock price of the company and Col-6 shows the ratio of the difference between maximum energy and mean energy to the standard deviation of energy. Abbreviation of technical setup:  SR=Support-Resistance, AC=Ascending-Channel, AT=Ascending Triangle, VCP=Volatility Contraction Pattern, AC=Ascending Channel, DB=Double Bottom}
\label{tab:2}
\tiny
\begin{tabular}{|l|l|l|l|l|r|}
\hline
\textbf{Sl. no.} &\textbf{COMPANY} & \textbf{Time-}  & \centering\textbf{Event}&\textbf{Impact on} & \textbf{Ratio} \\
&\textbf{NAME}& \textbf{scale}  &  &\textbf{stock price}&  \\
\cline{1-6}
 
1& GODREJCP& 10 min	&	Result			&Positive				&9.3\\
\hline

2& Ishanch  &10 min				&	Result			&Negative				&11.9\\
\hline

3& NINE&	Daily	&	Increase in earning			&Positive				&9.2\\
\hline

4& PNBHOUSING& Weekly		&	Carlyle Group to acquire controlling stake~\cite{pnb}  & Positive				&7.9\\
\hline

5& JUBLINGREA& Daily		& Technical setup(AC)				&Negative				&4.3\\
\hline

6& Boeing Co. & 5 min				&Technical setup	(DB)			&		Positive		&4.9\\
\hline

7& NK Rosneft' PAO& Daily & Russia-Ukraine War & Negative &10.6\\
\hline

8& Chevron Corp. & Weekly	& Russia-Ukraine War~\cite{chevron}	&	Positive  &5.0\\
\hline

9& Jocil Ltd.& Daily		& Technical setup(SR)				&Positive				&6.7\\
\hline

10& Avast plc & Daily		&	Merger		&Positive 	&8.5\\
\hline

11& 9939.HK & Daily	&	COVID-19 drug fails clinical trial			&	Negative & 6.5\\
\hline

12& BioNTech SE& Weekly		&Strong sale of COVID vaccine			&Positive				&4.3\\
\hline

13& Shopify Inc & 15 min	&Growth in	Performance			&Positive				&4.5\\
\hline

14& PHIA & 15 min				&	Inspection in its facilities		&Negative				&8.6\\
\hline

15& GameStop Corp.& Weekly	&	Manipulation	~\cite{gamestop}		&	Positive		&7.8\\
\hline

16& Netflix Inc &	Daily			&	Lost 2 lakh subscribers in Q1~\cite{netflx}		&	Negative		&7.7 \\
\hline

17& NVIDIA Corp.& Weekly				&	Technical setup (AC)			&		Positive		&5.8\\
\hline

18& Macy's Inc.& 15 min				& Technical setup	(AT)			&		Positive		&4.1\\
\hline

19& AT\&T Inc.& Daily				&Technical setup	(AT)			&		Positive		&4.9\\
\hline

20& Pinduoduo Inc.& Daily				&	Positive statement from China State Council~\cite{pinduo}			&		Positive		&5.7\\
\hline

21& SELAN & Daily		&	Antelopus Energy to buy stake	&	 Positive			&7.8 \\
\hline

22& ZEEL & Daily	&	Removal of MD			& Positive				& 6.0\\
\hline

23& ZEEL & Daily	&	Merger with Sony			&Positive				& 5.7\\
\hline

24& IRCTC Ltd.& 30 min	&	Overvaluation			&	Negative			&6.7\\
\hline

25& OXY& Weekly  &	Russia-Ukraine War			& Positive	&4.7\\
\hline

26& GLENMARK &	Daily		&	Enters 3rd clinical trial for COVID drug	&	Positive			&10.7\\
\hline

27& HDFCLIFE&	10 min	& Standard Life to sell stake  & Negative		&6.1\\
\hline

28& HAPPSTMNDS& Daily	&Buy recommendation	&	Positive	&5.3\\
\hline

29& TEXMOPIPES& Weekly		&	Technical Setup (AC)			& Positive				&8.3\\
\hline

 30& HITECHCORP&	Daily 	&	Technical Setup (AT)			&	Positive			&7.6\\
\hline

31& ADANIPORTS&	10 min			&	 NSDL freezes 3 FPI accounts			&	Negative			&9.5\\
\hline

32& Coal India Ltd.&	Daily			&	Russia-Ukraine War			&		Positive		&4.9\\
\hline

33& ALKYLAMINE&	Daily			&	Technical setup(AT)			&		Positive		&8.3\\
\hline

34& Bajaj Finance Ltd.& 15 min				&Technical setup (VCP)			&		Positive		&4.9\\
\hline

\end{tabular}
\end{table*}

Finally we have presented the forecast for stock prices up to two steps using machine learning method. Details are given below.
\subsection{Prediction of stock price during EE}
\label{sec:pred}

\begin{figure}[hbt!]
\includegraphics[angle=0, width=9cm, height=7.5cm]{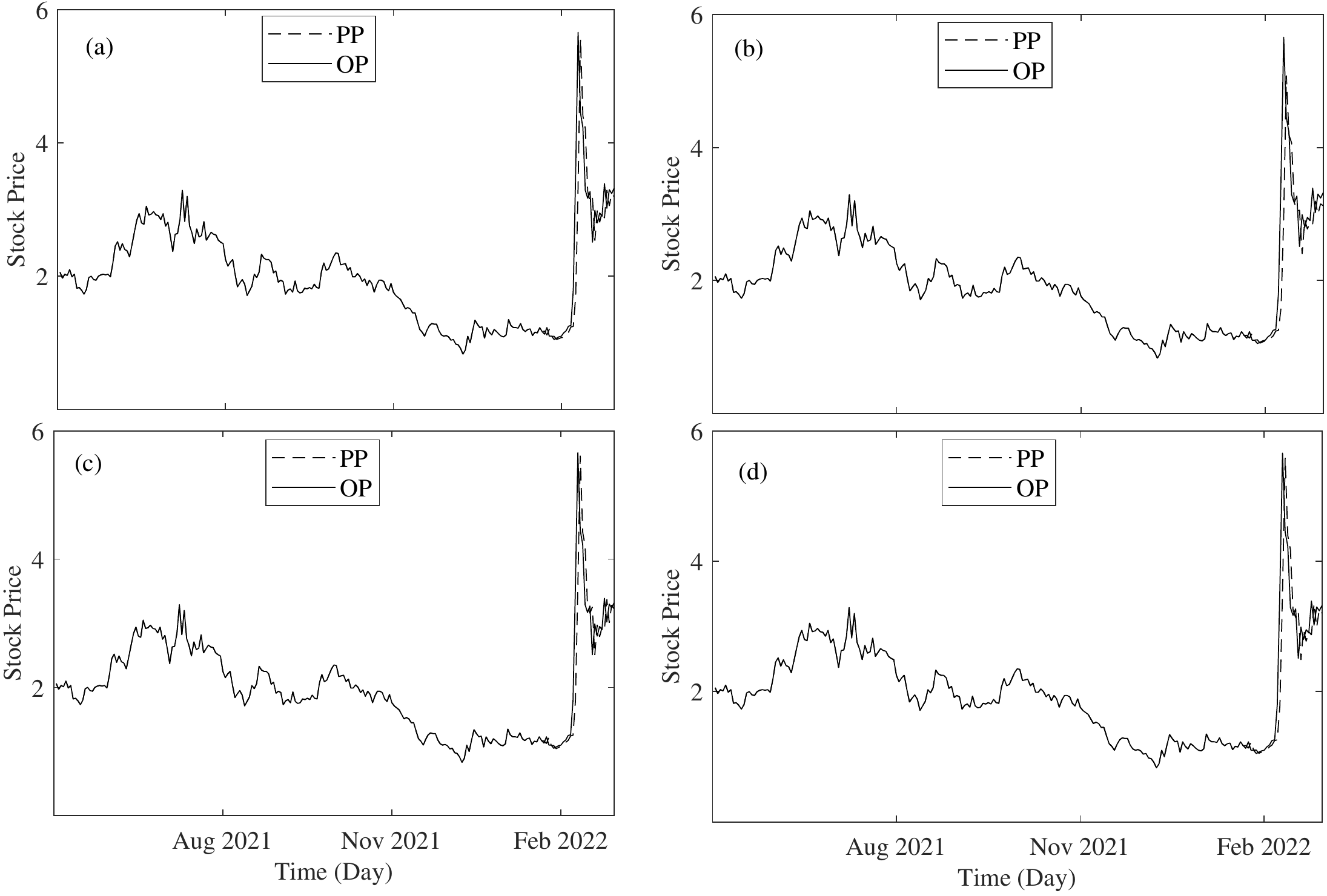}
\caption{\label{fig:predOD}(a) \& (c) represent the original stock price (OP) and 1-step ahead predicted stock price (PP) using OHLC price and only close price, respectively. Similarly, (b) \& (d) represent the original stock price (OP) and 2-step ahead predicted stock price (PP) using OHLC price and only close price, respectively, of Nine Energy Service Inc. during positive EE.}
\end{figure}

Using the SVR method, we conducted 1-step and 2-step ahead prediction of stock prices based on the previous OHLC (Open, High, Low, and Close) prices and only close price, respectively.

\begin{figure}[hbt!]
\includegraphics[angle=0, width=8cm, height=7.5cm]{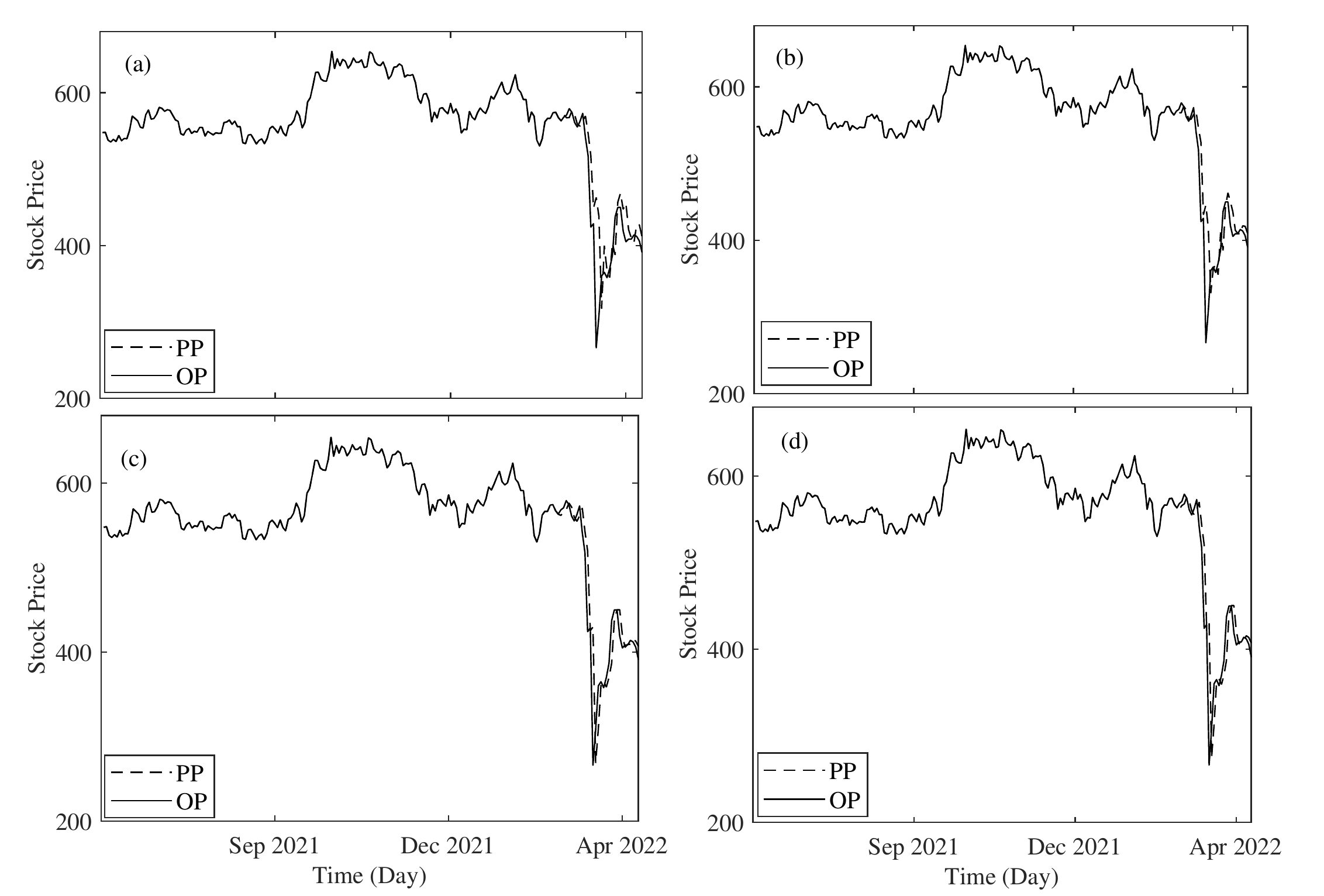}
\caption{\label{fig:predRosOD}(a) \& (c) represent the original stock price (OP) and 1-step ahead predicted stock price (PP) using OHLC price and only close price, respectively. Similarly, (b) \& (d) represent the original stock price (OP) and 2-step ahead predicted stock price (PP) using OHLC price and only close price, respectively, of NK Rosneft' PAO during negative EE.}
\end{figure}

Firstly, we predicted 1-step ahead using the inputs open-high-low-close (OHLC) prices. Fig.~\ref{fig:predOD}(a) and (c) show the original stock price (OP) and the one-step ahead predicted stock price (PP) of Nine Energy Service using OHLC price and only close price, respectively. Table~\ref{tab:3} shows the 1-step ahead prediction results for rest of the companies with their mean absolute percentage error (MAPE) and accuracy. 30-blocks represents the prediction result for 30 steps when EE had occurred. 5-blocks show the prediction result for 5 steps and within these 5 steps EE had occurred and 1-block shows the stock price prediction result on  EE. Fig.~\ref{fig:predOD} shows the prediction of stock price during positive EE and, Fig.~\ref{fig:predRosOD} shows the stock price prediction during negative EE. The average MAPE for 30-blocks, 5-blocks, 1-block using OHLC price as input are 4.32\%, 9.82\% and 23.33\%, respectively. Whereas using  only close price (C) as input the average MAPEs are 4.02\%, 9.13\% and 22.48\%, respectively. According to the findings, the closer we try to predict the stock price to the EE, the lower the accuracy of the prediction. This indicates that predicting the stock price specifically during the EE is challenging. 

\begin{figure}[hbt!]
\includegraphics[angle=0, width=7cm, height=6cm]{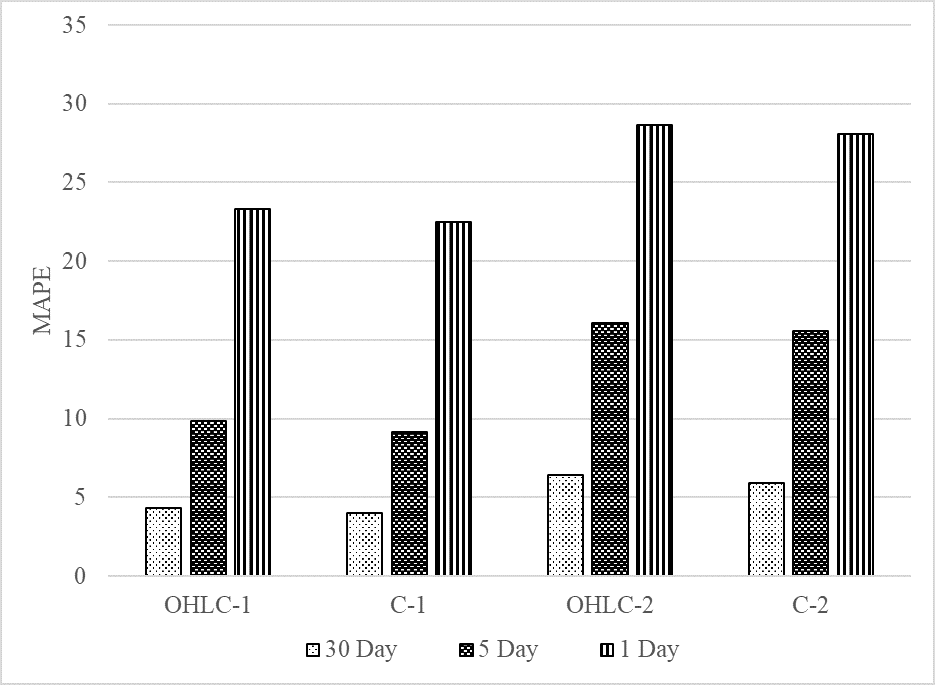}
\caption{\label{fig:chart}Plot represents the average MAPE for 1-step ahead prediction using open-high-low-close price (OHLC-1), only close (C-1) and 2-step ahead prediction using open-high-low-close price (OHLC-2) and only close price (C-2)}
\end{figure}

Similar results are obtained using 2-step ahead prediction which is shown in Table~\ref{tab:4}. The average MAPE for 30-blocks, 5-blocks and 1-block using OHLC prices as input are 6.42\%, 16.06\% and 28.64\%, respectively. Whereas using only the close price the average MAPEs are 5.91\%, 15.55\% and 28.09\%, respectively. The average MAPE for 1-step ahead and 2-step ahead prediction using only close price, and using OHLC price, as input is shown in Fig.~\ref{fig:chart}. It is clear form the table~\ref{tab:3}, table~\ref{tab:4} and Fig.~\ref{fig:chart} that the 1-step ahead prediction is more accurate than the 2-step ahead prediction. Furthermore, utilizing solely the closing price as an input feature marginally enhances the precision of the prediction compared to utilizing OHLC prices as input.

\begin{table*}[!ht]
\caption{Table represents the MAPE and accuracy of 1-step ahead prediction using SVR. Col-1 represents the companies that are analyzed. Col-2 represents the 1-step ahead prediction using open-high-low-close prices (OHLC) as input feature. Col-3 represents the 1-step ahead prediction using close price (C) as input feature. The sub-columns in Col-2 and Col-3 show the MAPE and accuracy of 30-blocks, 5-blocks and 1-block prediction when an EE had occurred.}
\label{tab:3}
\tiny
\begin{tabular}{|l|r|r|r|r|r|r|r|r|r|r|r|r|}
\hline
 &\multicolumn{6}{| c |}{1-step ahead prediction (OHLC)}  & \multicolumn{6}{| c |}{1-step ahead prediction (C)}\\
\cline{2-13}
\textbf{COMPANY} &\multicolumn{2}{| c |}{30-blocks}  & \multicolumn{2}{| c |}{5-blocks}& \multicolumn{2}{| c |}{1-block}&\multicolumn{2}{| c |}{30-blocks}  & \multicolumn{2}{| c |}{5-blocks}& \multicolumn{2}{| c |}{1-block} \\
\cline{2-13}
 & MAPE&Accuracy& MAPE&Accuracy& MAPE&Accuracy& MAPE&Accuracy& MAPE&Accuracy& MAPE&Accuracy\\
\hline 
 ADANIPORTS&4.32 &95.68&8.20 &91.80&13.84 &86.16& 2.19&97.81&4.42 &95.58&13.90 &86.10\\
\hline
 SELAN&2.81 &97.19&6.22 &93.78&17.04 &82.96&2.46&97.54 &5.41&94.59 &16.90&83.10 \\
\hline
Coal India Ltd. & 2.51&97.49&6.20 &93.80&8.85 &91.15&2.45 &97.55&5.69 &94.31&8.51 &91.49\\
\hline
 Ishanch& 1.81&98.19&6.02 &93.98&18.36 &81.64&1.77 &98.23&4.72 &95.28&18.39 &81.61\\
\hline
 NINE&11.13 &88.87&30.25 &69.75&52.70 &47.30&11.03 &88.97&28.97 &71.03&50.08 &49.92\\
\hline
NK Rosneft' PAO &8.35 &91.65&30.99 &69.01&73.90 &26.10&5.89 &94.11&22.32 &77.68&61.39 &38.91\\
\hline
 GLENMARK&2.42 &97.58&5.84 &94.16&21.51 &78.49&2.64 &97.36&7.41 &92.59&21.39 &78.61\\
\hline
 Jocil Ltd.&2.69 &97.31&4.76 &95.24&16.71 &83.29&2.60 &97.40&10.54 &89.46&16.60 &83.40\\
\hline
 AT\&T Inc.&1.14 &98.86&2.03 &97.97&3.67 &96.33&1.12 &98.88&1.91 &98.09&3.74 &96.26\\
\hline
 HDFC LIFE&0.24 &99.76&0.72 &99.28& 1.70&98.30&0.22 &99.78&0.62 &99.38&1.62 &98.38\\
\hline
 HITECHCORP&2.75 &97.25&6.01 &93.99&16.60 &83.40&3.03 &96.97&7.75 &92.25& 16.56&83.44\\
\hline
 ALKYLAMINE&3.33 &96.67&9.51 &90.49&15.10 &84.90&5.02 &94.98&9.48 &90.52&17.41 &82.59\\
\hline
 Chevron Corp.&3.26 &96.72&3.59 &96.41&11.86 &88.14&3.46 &96.54&6.41 &93.59&12.29 &87.71\\
\hline
 BioNTech SE&4.25 &95.75&7.14 &92.86&11.54 &88.46&4.13 &95.87&5.62 &94.38&15.06 &84.94\\
\hline
 Netflix Inc& 5.88&94.12&18.43 &81.57&60.14 &39.86&4.64 &95.36&13.49 &86.51&55.29 &44.71\\
\hline
 NVIDIA Corp.&6.45 &93.55&7.76 &92.24&18.41 &81.59&4.96 &95.04&7.47 &92.53&15.65 &84.35\\
\hline
 Pinduoduo Inc.&5.14 &94.86&6.71 &93.29&13.60 &86.4&5.08 &94.92&7.70 &92.30&13.26 &86.74\\
\hline
 OXY&6.66 &93.34&9.24 &90.76&32.54 &67.46&6.46 &93.54&9.39 &90.61&31.59 & 68.41\\
\hline
 TEXMOPIPES&6.70 &93.30&15.52 &84.48&36.76 &63.24&6.82 &93.18&14.02 &85.98&37.47 &62.53\\
\hline
 PNBHOUSING&4.53 &95.47&11.30 &88.70&21.60 &78.40&4.38 &95.62& 9.62&90.38&17.32 &82.68\\
\hline
\textbf{Mean} &\textbf{4.32} &\textbf{95.64}&\textbf{9.82} &\textbf{90.18}&\textbf{23.33} &\textbf{76.67}&\textbf{4.02} &\textbf{95.98}&\textbf{9.13} &\textbf{90.87}&\textbf{22.48} &\textbf{77.52}\\
\hline
\end{tabular}
\end{table*}

\begin{table*}[!ht]
\caption{Table represents the MAPE and accuracy of 2-step ahead prediction using SVR. Col-1 represents the companies that are analyzed. Col-2 represents the 2-step ahead prediction using OHLC prices as input feature. Col-3 represents the 2-step ahead prediction using close price as input feature. The sub-columns in Col-2 and Col-3 shows the MAPE and accuracy of 30-blocks, 5-blocks and 1-block prediction when an EE had occurred.}
\label{tab:4}
\tiny
\begin{tabular}{|l|r|r|r|r|r|r|r|r|r|r|r|r|}
\hline
\textbf{COMPANY} &\multicolumn{6}{| c |}{2-step ahead prediction(OHLC)}  & \multicolumn{6}{| c |}{2-step ahead prediction(C)}\\
\cline{2-13}
 &\multicolumn{2}{| c |}{30-blocks}  & \multicolumn{2}{| c |}{5-blocks}& \multicolumn{2}{| c |}{1-block}&\multicolumn{2}{| c |}{30-blocks}  & \multicolumn{2}{| c |}{5-blocks}& \multicolumn{2}{| c |}{1-block} \\
\cline{2-13}
 & MAPE&Accuracy& MAPE&Accuracy& MAPE&Accuracy& MAPE&Accuracy& MAPE&Accuracy& MAPE&Accuracy\\
\hline 
 ADANIPORTS&6.60 &93.40&12.61 &87.39&17.27 &82.73&2.97 &97.03&9.54 &90.46&17.34 &82.66\\
\hline
 SELAN&4.52 &95.48 &11.45 &88.55&19.84 &80.16&4.46&95.54 &11.22&88.78 &19.81&80.19\\
\hline
Coal India Ltd. &3.17 &96.83&7.99 &92.01&12.98 &87.02&3.23 &96.77&7.88 &92.12&11.98 &88.02\\
\hline
 Ishanch&2.85 &97.15&12.01 &87.99&22.34 &77.66&2.57 &97.43&10.17 &89.83&22.19 &77.81\\
\hline
 NINE&14.43 &85.57&44.41 &55.59&71.85 &28.15&14.32 &85.68&43.24 &56.76&68.39 &31.61\\
\hline
NK Rosneft' PAO &9.63 &90.37&34.95 &65.05&62.69 &37.31&9.77 &90.23 &34.66 &65.34&60.09 &39.91\\
\hline
 GLENMARK&3.65 &96.35&10.08 &89.92&22.04 &77.96&4.08 &95.92&11.83 &88.17&22.11 &77.89\\
\hline
 Jocil Ltd.&4.29 &95.71&16.46 &83.54&30.15 &69.85&4.37 &95.63&16.42 &83.58&30.02 &69.98\\
\hline
 AT\&T Inc.&1.69 &98.31&3.08 &96.92&5.77 &94.23&1.66 &98.34&2.30 &97.70&5.27 &94.73\\
\hline
 HDFC LIFE&0.37 &99.63&1.09 &98.91&2.65 &97.35&0.39 &99.61&1.16 &98.84&2.81 &97.19\\
\hline
 HITECHCORP&4.66 &95.34&13.11 &86.89&24.21 &75.79&4.63 &95.37&13.76 &86.24&24.02 &75.98\\
\hline
 ALKYLAMINE&6.19 &93.81&10.98 &89.02&20.28 &79.72&4.08 &95.92&12.71 &87.29&23.57 &76.43\\
\hline
 Chevron Corp.&4.31 &95.69&8.01 &91.99&18.42 &81.58&4.63 &95.37&8.49 &91.51&18.82 &81.18\\
\hline
 BioNTech SE&6.72 &93.28&10.59 &89.41&16.73 &83.27&7.11 &92.89&10.72 &89.28&17.87 &82.13\\
\hline
 Netflix Inc&9.47 &90.53&28.12 &71.88&63.04 &36.96&7.39 &92.61& 27.91&72.09&61.48 &38.52\\
\hline
 NVIDIA Corp.&8.04 &91.96&12.46 &87.54&22.01 &77.99&8.12 &91.88&12.38 &87.62&21.79 &78.21\\
\hline
 Pinduoduo Inc.&8.18 &91.82&12.41 &87.59&23.29 &76.71&7.28 &92.72& 12.59&87.41&20.14 &79.86\\
\hline
 OXY&12.12 &87.88&23.91 &76.09&38.47 &61.53&9.71 &90.29&17.89 &82.11&33.68 &66.32\\
\hline
 TEXMOPIPES&9.79 &90.21&23.94 &76.06&47.53 &52.47&10.15 &89.85&23.76 &76.24&49.08 &50.92\\
\hline
 PNBHOUSING&7.72 &92.28& 23.50&76.50&31.20 &68.80&7.28 &92.72&22.34 &77.66&31.38 &68.62\\
\hline
 \textbf{Mean}& \textbf{6.42}&\textbf{93.58}&\textbf{16.06} &\textbf{83.94}& \textbf{28.64}&\textbf{71.36}&\textbf{5.91} &\textbf{94.09}&\textbf{15.55} &\textbf{84.45}&\textbf{28.09} &\textbf{71.91}\\
\hline
\end{tabular}
\end{table*}

\begin{figure}[!ht]
\includegraphics[angle=0, width=8cm, height=7.5cm]{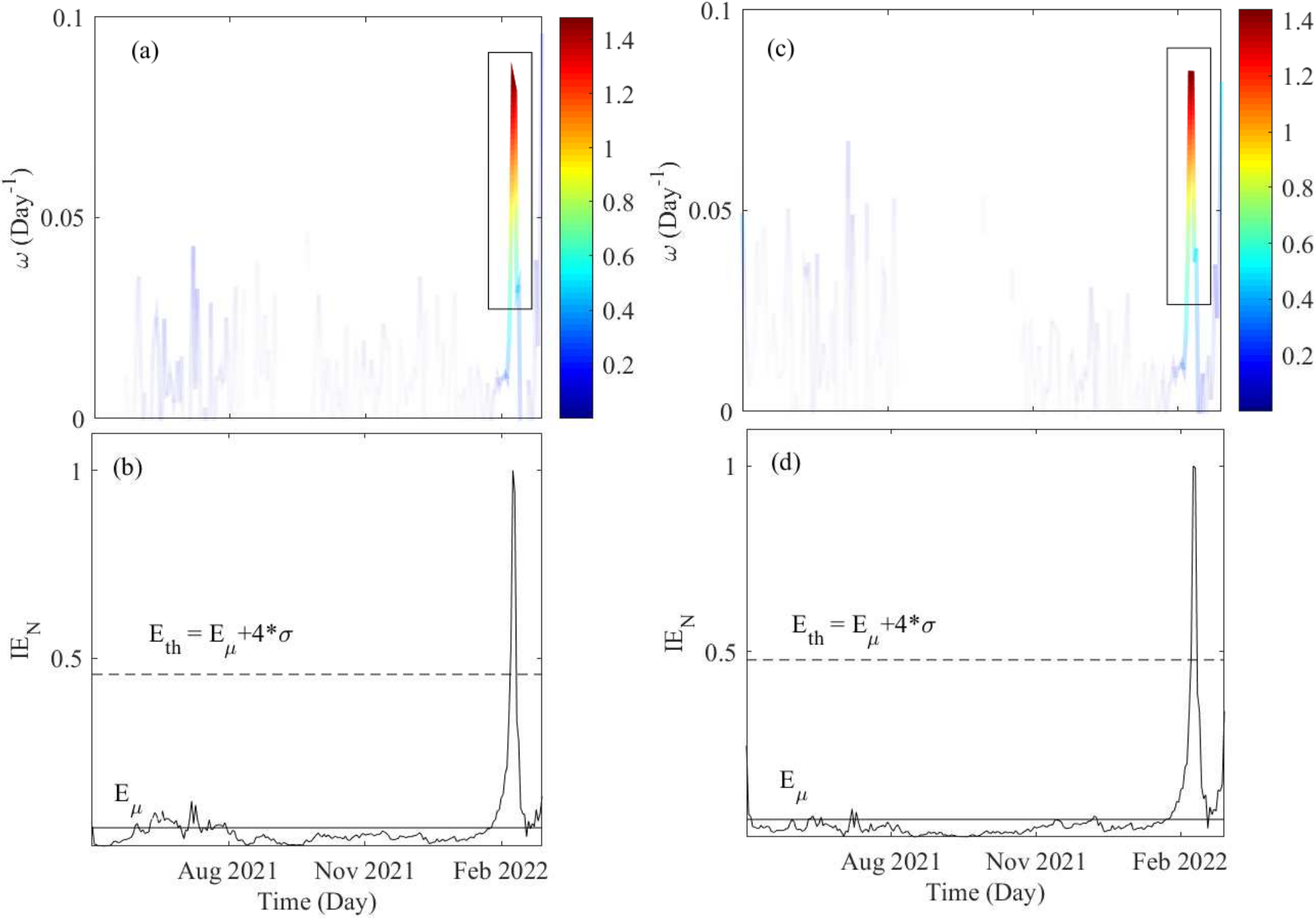}
\caption{\label{fig:Nine_1}Plot (a) \& (b) represent the $H(t,\omega)$ and $IE_N(t)$ plot of  1-step ahead predicted stock price using OHLC price, and plot (c) \& (d) represent the $H(t,\omega)$ and $IE_N(t)$ plot of 1-step ahead predicted stock price using only close price of Nine Energy Service Inc., respectively.}
\end{figure}

\begin{figure}[hbt!]
\includegraphics[angle=0, width=9cm, height=7.5cm]{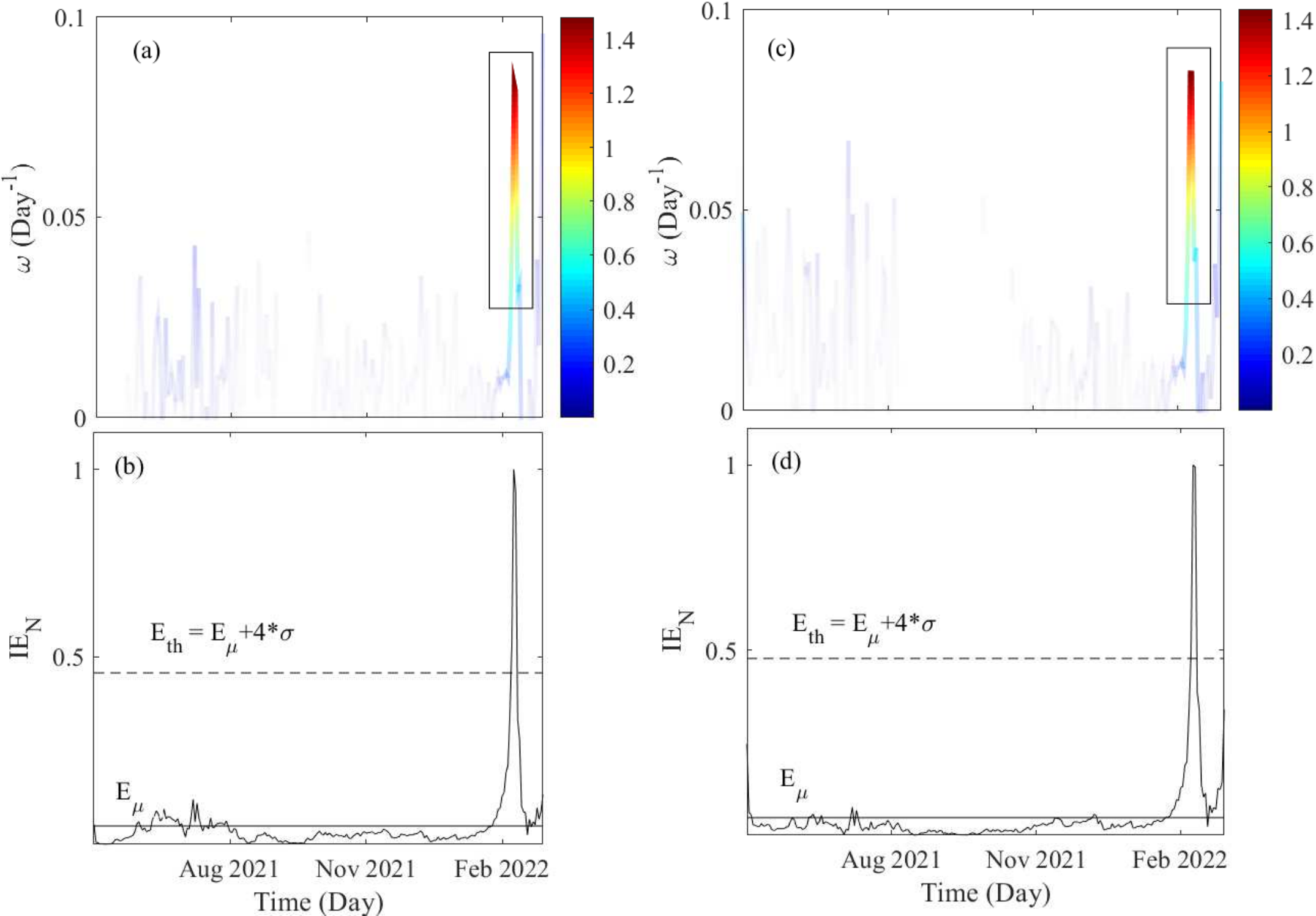}
\caption{\label{fig:Nine_2}Plot (a) \& (b) represent the $H(t,\omega)$ and $IE_N(t)$ plot of  2-step ahead predicted stock price using OHLC price, and plot (c) \& (d) represent the $H(t,\omega)$ and $IE_N(t)$ plot of 2-step ahead predicted stock price using only close price of Nine Energy Service Inc., respectively.}
\end{figure}

Using the one-step ahead predicted time-series obtained from the SVR, we have identified both the positive and negative EE. The $H(t,\omega)$ of Nine Energy Service Inc. using predicted time-series obtained from SVR with OHLC prices and only close price, respectively, are shown in Fig.~\ref{fig:Nine_1} (a) \& (c). The EE is captured by a high energy concentration which is shown by the red patch in Fig.~\ref{fig:Nine_1} (a) \& (c). Further, $IE_N(t)$ in Fig.~\ref{fig:Nine_1} (b) \& (d) clearly shows a spike which is greater than the $E_{th}$. As the $IE_N$ during the stock price crash is greater than $E_{th}$, the event can be considered as a negative EE. EE is also identified from the other time-series obtained from the SVR. The identification of positive and negative EE was accomplished by utilizing the two-step ahead predicted time-series. Fig.~\ref{fig:Nine_2} demonstrates that these time-series effectively capture a high energy concentration, with a spike in $IE_N(t)$ that exceeds $E_{th}$, thereby identifying EE in the stock price. These outcomes serve as compelling evidence that the predicted time-series possess the capability to forecast EE.

\section{Conclusion}
\label{sec:con}

This study investigated the factors responsible for inducing sudden positive or negative extreme events (EEs) in the stock price from a normal state. The results indicate that these EEs can be attributed to fundamental, technical, and external factors. Specifically, positive EEs were primarily associated with exceptional financial results, anticipated future growth, and acquisition of stakes, while negative EEs were found to occur in response to negative fundamental factors, such as poor financial results. Additionally, technical factors, such as bullish chart patterns like double bottom, were associated with positive EEs, whereas bearish chart patterns like channel formation were linked to negative EEs. Finally, external factors, such as wars, were found to influence the occurrence of both positive and negative EEs in the stock market. Positive EEs were observed in companies that maintained a positive business outlook during times of war, while negative EEs were identified in companies whose business outlook was negatively affected by the war.

We have determined that, for extreme events (EEs) with both positive and negative outcomes, the instantaneous energy ($IE_N$) exceeds a specific threshold of $E_{\mu}+4\sigma$. Here, $E_{\mu}$ and $\sigma$ represent the mean and standard deviation of energy, respectively. This condition holds true for all three factors involved in the EEs. We applied Support Vector Regression (SVR) to forecast stock prices during an EE in the stock market. We conducted 1-step ahead and 2-step ahead predictions using both the close price and OHLC prices as inputs. The results indicate that the 1-step prediction is more accurate than the 2-step prediction. Additionally, using only the close price as input slightly increases the accuracy compared to using the OHLC price. Accuracy decreases when predicting stock prices closer to an EE, with the least accuracy occurring at the time of the EE itself. This demonstrates that forecasting stock prices during an EE is extremely challenging. However, the predicted time-series from the SVR was used to identify the EE, and the results demonstrate that the predicted time-series can successfully identify the EE. Therefore, we can conclude that the SVR can moderately predict stock prices during an EE.

Detecting EEs resulting from various factors is crucial, as it can assist investors and traders in making informed investment decisions. Although it is challenging to make a profit from positive EEs, one can leverage the opportunity to buy into a fundamentally strong stock at a lower price when the stock experiences a crash due to negative EEs triggered by technical and external factors. This analysis demonstrates that maintaining a cash position in a portfolio is crucial to capitalize on such opportunities. The error values indicate that predicting EE is a challenging task. However, in future studies, application of more advanced techniques can potentially reduce the error.

\section{Acknowledgements}
We express our gratitude to NIT Sikkim for providing doctoral fellowships to AR and SRL. We also extend our appreciation to Sandeep Parajuli for his valuable assistance in this research.

\bibliography{apssamp}% Produces the bibliography via BibTeX.

\begin{thebibliography}{63}%
\makeatletter
\providecommand \@ifxundefined [1]{%
 \@ifx{#1\undefined}
}%
\providecommand \@ifnum [1]{%
 \ifnum #1\expandafter \@firstoftwo
 \else \expandafter \@secondoftwo
 \fi
}%
\providecommand \@ifx [1]{%
 \ifx #1\expandafter \@firstoftwo
 \else \expandafter \@secondoftwo
 \fi
}%
\providecommand \natexlab [1]{#1}%
\providecommand \enquote  [1]{``#1''}%
\providecommand \bibnamefont  [1]{#1}%
\providecommand \bibfnamefont [1]{#1}%
\providecommand \citenamefont [1]{#1}%
\providecommand \href@noop [0]{\@secondoftwo}%
\providecommand \href [0]{\begingroup \@sanitize@url \@href}%
\providecommand \@href[1]{\@@startlink{#1}\@@href}%
\providecommand \@@href[1]{\endgroup#1\@@endlink}%
\providecommand \@sanitize@url [0]{\catcode `\\12\catcode `\$12\catcode
  `\&12\catcode `\#12\catcode `\^12\catcode `\_12\catcode `\%12\relax}%
\providecommand \@@startlink[1]{}%
\providecommand \@@endlink[0]{}%
\providecommand \url  [0]{\begingroup\@sanitize@url \@url }%
\providecommand \@url [1]{\endgroup\@href {#1}{\urlprefix }}%
\providecommand \urlprefix  [0]{URL }%
\providecommand \Eprint [0]{\href }%
\providecommand \doibase [0]{http://dx.doi.org/}%
\providecommand \selectlanguage [0]{\@gobble}%
\providecommand \bibinfo  [0]{\@secondoftwo}%
\providecommand \bibfield  [0]{\@secondoftwo}%
\providecommand \translation [1]{[#1]}%
\providecommand \BibitemOpen [0]{}%
\providecommand \bibitemStop [0]{}%
\providecommand \bibitemNoStop [0]{.\EOS\space}%
\providecommand \EOS [0]{\spacefactor3000\relax}%
\providecommand \BibitemShut  [1]{\csname bibitem#1\endcsname}%
\let\auto@bib@innerbib\@empty
%</preamble>
\bibitem [{\citenamefont {Bunde}\ \emph {et~al.}(2012)\citenamefont {Bunde},
  \citenamefont {Kropp},\ and\ \citenamefont
  {Schellnhuber}}]{bunde2012science}%
  \BibitemOpen
  \bibfield  {author} {\bibinfo {author} {\bibfnamefont {A.}~\bibnamefont
  {Bunde}}, \bibinfo {author} {\bibfnamefont {J.}~\bibnamefont {Kropp}}, \ and\
  \bibinfo {author} {\bibfnamefont {H.-J.}\ \bibnamefont {Schellnhuber}},\
  }\href@noop {} {\emph {\bibinfo {title} {The science of disasters: climate
  disruptions, heart attacks, and market crashes}}}\ (\bibinfo  {publisher}
  {Springer Science \& Business Media},\ \bibinfo {year} {2012})\BibitemShut
  {NoStop}%
\bibitem [{\citenamefont {Zhang}\ \emph {et~al.}(2009)\citenamefont {Zhang},
  \citenamefont {Yu}, \citenamefont {Wang},\ and\ \citenamefont
  {Lai}}]{zhang2009estimating}%
  \BibitemOpen
  \bibfield  {author} {\bibinfo {author} {\bibfnamefont {X.}~\bibnamefont
  {Zhang}}, \bibinfo {author} {\bibfnamefont {L.}~\bibnamefont {Yu}}, \bibinfo
  {author} {\bibfnamefont {S.}~\bibnamefont {Wang}}, \ and\ \bibinfo {author}
  {\bibfnamefont {K.~K.}\ \bibnamefont {Lai}},\ }\href@noop {} {\bibfield
  {journal} {\bibinfo  {journal} {Energy Economics}\ }\textbf {\bibinfo
  {volume} {31}},\ \bibinfo {pages} {768} (\bibinfo {year} {2009})}\BibitemShut
  {NoStop}%
\bibitem [{\citenamefont {Albeverio}\ \emph {et~al.}(2006)\citenamefont
  {Albeverio}, \citenamefont {Jentsch},\ and\ \citenamefont
  {Kantz}}]{albeverio2006extreme}%
  \BibitemOpen
  \bibfield  {author} {\bibinfo {author} {\bibfnamefont {S.}~\bibnamefont
  {Albeverio}}, \bibinfo {author} {\bibfnamefont {V.}~\bibnamefont {Jentsch}},
  \ and\ \bibinfo {author} {\bibfnamefont {H.}~\bibnamefont {Kantz}},\
  }\href@noop {} {\emph {\bibinfo {title} {Extreme events in nature and
  society}}}\ (\bibinfo  {publisher} {Springer Science \& Business Media},\
  \bibinfo {year} {2006})\BibitemShut {NoStop}%
\bibitem [{\citenamefont {Mahata}\ \emph
  {et~al.}(2021{\natexlab{a}})\citenamefont {Mahata}, \citenamefont {Rai},
  \citenamefont {Nurujjaman}, \citenamefont {Prakash},\ and\ \citenamefont
  {Prasad~Bal}}]{mahata2021characteristics}%
  \BibitemOpen
  \bibfield  {author} {\bibinfo {author} {\bibfnamefont {A.}~\bibnamefont
  {Mahata}}, \bibinfo {author} {\bibfnamefont {A.}~\bibnamefont {Rai}},
  \bibinfo {author} {\bibfnamefont {M.}~\bibnamefont {Nurujjaman}}, \bibinfo
  {author} {\bibfnamefont {O.}~\bibnamefont {Prakash}}, \ and\ \bibinfo
  {author} {\bibfnamefont {D.}~\bibnamefont {Prasad~Bal}},\ }\href@noop {}
  {\bibfield  {journal} {\bibinfo  {journal} {Chaos: An Interdisciplinary
  Journal of Nonlinear Science}\ }\textbf {\bibinfo {volume} {31}},\ \bibinfo
  {pages} {053115} (\bibinfo {year} {2021}{\natexlab{a}})}\BibitemShut
  {NoStop}%
\bibitem [{\citenamefont {Banerjee}\ \emph {et~al.}()\citenamefont {Banerjee},
  \citenamefont {Mishra}, \citenamefont {Dana}, \citenamefont {Hens},
  \citenamefont {Kapitaniak}, \citenamefont {Kurths},\ and\ \citenamefont
  {Marwan}}]{banerjeepredicting}%
  \BibitemOpen
  \bibfield  {author} {\bibinfo {author} {\bibfnamefont {A.}~\bibnamefont
  {Banerjee}}, \bibinfo {author} {\bibfnamefont {A.}~\bibnamefont {Mishra}},
  \bibinfo {author} {\bibfnamefont {S.~K.}\ \bibnamefont {Dana}}, \bibinfo
  {author} {\bibfnamefont {C.}~\bibnamefont {Hens}}, \bibinfo {author}
  {\bibfnamefont {T.}~\bibnamefont {Kapitaniak}}, \bibinfo {author}
  {\bibfnamefont {J.}~\bibnamefont {Kurths}}, \ and\ \bibinfo {author}
  {\bibfnamefont {N.}~\bibnamefont {Marwan}},\ }\href@noop {} {\bibinfo
  {journal} {Frontiers in Applied Mathematics and Statistics}\ ,\ \bibinfo
  {pages} {99}}\BibitemShut {NoStop}%
\bibitem [{\citenamefont {Mishra}\ \emph {et~al.}(2018)\citenamefont {Mishra},
  \citenamefont {Saha}, \citenamefont {Vigneshwaran}, \citenamefont {Pal},
  \citenamefont {Kapitaniak},\ and\ \citenamefont {Dana}}]{mishra2018dragon}%
  \BibitemOpen
\bibfield  {journal} {  }\bibfield  {author} {\bibinfo {author} {\bibfnamefont
  {A.}~\bibnamefont {Mishra}}, \bibinfo {author} {\bibfnamefont
  {S.}~\bibnamefont {Saha}}, \bibinfo {author} {\bibfnamefont {M.}~\bibnamefont
  {Vigneshwaran}}, \bibinfo {author} {\bibfnamefont {P.}~\bibnamefont {Pal}},
  \bibinfo {author} {\bibfnamefont {T.}~\bibnamefont {Kapitaniak}}, \ and\
  \bibinfo {author} {\bibfnamefont {S.~K.}\ \bibnamefont {Dana}},\ }\href@noop
  {} {\bibfield  {journal} {\bibinfo  {journal} {Physical Review E}\ }\textbf
  {\bibinfo {volume} {97}},\ \bibinfo {pages} {062311} (\bibinfo {year}
  {2018})}\BibitemShut {NoStop}%
\bibitem [{\citenamefont {Karnatak}\ \emph {et~al.}(2014)\citenamefont
  {Karnatak}, \citenamefont {Ansmann}, \citenamefont {Feudel},\ and\
  \citenamefont {Lehnertz}}]{karnatak2014route}%
  \BibitemOpen
  \bibfield  {author} {\bibinfo {author} {\bibfnamefont {R.}~\bibnamefont
  {Karnatak}}, \bibinfo {author} {\bibfnamefont {G.}~\bibnamefont {Ansmann}},
  \bibinfo {author} {\bibfnamefont {U.}~\bibnamefont {Feudel}}, \ and\ \bibinfo
  {author} {\bibfnamefont {K.}~\bibnamefont {Lehnertz}},\ }\href@noop {}
  {\bibfield  {journal} {\bibinfo  {journal} {Physical Review E}\ }\textbf
  {\bibinfo {volume} {90}},\ \bibinfo {pages} {022917} (\bibinfo {year}
  {2014})}\BibitemShut {NoStop}%
\bibitem [{\citenamefont {Malik}\ and\ \citenamefont
  {Ozturk}(2020)}]{malik2020rare}%
  \BibitemOpen
  \bibfield  {author} {\bibinfo {author} {\bibfnamefont {N.}~\bibnamefont
  {Malik}}\ and\ \bibinfo {author} {\bibfnamefont {U.}~\bibnamefont {Ozturk}},\
  }\href@noop {} {\bibfield  {journal} {\bibinfo  {journal} {Chaos: An
  Interdisciplinary Journal of Nonlinear Science}\ }\textbf {\bibinfo {volume}
  {30}},\ \bibinfo {pages} {090401} (\bibinfo {year} {2020})}\BibitemShut
  {NoStop}%
\bibitem [{\citenamefont {Chowdhury}\ \emph {et~al.}(2022)\citenamefont
  {Chowdhury}, \citenamefont {Ray}, \citenamefont {Dana},\ and\ \citenamefont
  {Ghosh}}]{chowdhury2022extreme}%
  \BibitemOpen
  \bibfield  {author} {\bibinfo {author} {\bibfnamefont {S.~N.}\ \bibnamefont
  {Chowdhury}}, \bibinfo {author} {\bibfnamefont {A.}~\bibnamefont {Ray}},
  \bibinfo {author} {\bibfnamefont {S.~K.}\ \bibnamefont {Dana}}, \ and\
  \bibinfo {author} {\bibfnamefont {D.}~\bibnamefont {Ghosh}},\ }\href@noop {}
  {\bibfield  {journal} {\bibinfo  {journal} {Physics Reports}\ }\textbf
  {\bibinfo {volume} {966}},\ \bibinfo {pages} {1} (\bibinfo {year}
  {2022})}\BibitemShut {NoStop}%
\bibitem [{\citenamefont {Ray}\ \emph {et~al.}(2020)\citenamefont {Ray},
  \citenamefont {Mishra}, \citenamefont {Ghosh}, \citenamefont {Kapitaniak},
  \citenamefont {Dana},\ and\ \citenamefont {Hens}}]{ray2020extreme}%
  \BibitemOpen
  \bibfield  {author} {\bibinfo {author} {\bibfnamefont {A.}~\bibnamefont
  {Ray}}, \bibinfo {author} {\bibfnamefont {A.}~\bibnamefont {Mishra}},
  \bibinfo {author} {\bibfnamefont {D.}~\bibnamefont {Ghosh}}, \bibinfo
  {author} {\bibfnamefont {T.}~\bibnamefont {Kapitaniak}}, \bibinfo {author}
  {\bibfnamefont {S.~K.}\ \bibnamefont {Dana}}, \ and\ \bibinfo {author}
  {\bibfnamefont {C.}~\bibnamefont {Hens}},\ }\href@noop {} {\bibfield
  {journal} {\bibinfo  {journal} {Physical Review E}\ }\textbf {\bibinfo
  {volume} {101}},\ \bibinfo {pages} {032209} (\bibinfo {year}
  {2020})}\BibitemShut {NoStop}%
\bibitem [{\citenamefont {McPhillips}\ \emph {et~al.}(2018)\citenamefont
  {McPhillips}, \citenamefont {Chang}, \citenamefont {Chester}, \citenamefont
  {Depietri}, \citenamefont {Friedman}, \citenamefont {Grimm}, \citenamefont
  {Kominoski}, \citenamefont {McPhearson}, \citenamefont
  {M{\'e}ndez-L{\'a}zaro}, \citenamefont {Rosi} \emph
  {et~al.}}]{mcphillips2018defining}%
  \BibitemOpen
  \bibfield  {author} {\bibinfo {author} {\bibfnamefont {L.~E.}\ \bibnamefont
  {McPhillips}}, \bibinfo {author} {\bibfnamefont {H.}~\bibnamefont {Chang}},
  \bibinfo {author} {\bibfnamefont {M.~V.}\ \bibnamefont {Chester}}, \bibinfo
  {author} {\bibfnamefont {Y.}~\bibnamefont {Depietri}}, \bibinfo {author}
  {\bibfnamefont {E.}~\bibnamefont {Friedman}}, \bibinfo {author}
  {\bibfnamefont {N.~B.}\ \bibnamefont {Grimm}}, \bibinfo {author}
  {\bibfnamefont {J.~S.}\ \bibnamefont {Kominoski}}, \bibinfo {author}
  {\bibfnamefont {T.}~\bibnamefont {McPhearson}}, \bibinfo {author}
  {\bibfnamefont {P.}~\bibnamefont {M{\'e}ndez-L{\'a}zaro}}, \bibinfo {author}
  {\bibfnamefont {E.~J.}\ \bibnamefont {Rosi}},  \emph {et~al.},\ }\href@noop
  {} {\bibfield  {journal} {\bibinfo  {journal} {Earth's Future}\ }\textbf
  {\bibinfo {volume} {6}},\ \bibinfo {pages} {441} (\bibinfo {year}
  {2018})}\BibitemShut {NoStop}%
\bibitem [{\citenamefont {Rai}\ \emph {et~al.}(2022{\natexlab{a}})\citenamefont
  {Rai}, \citenamefont {Mahata}, \citenamefont {Nurujjaman},\ and\
  \citenamefont {Prakash}}]{rai2022statistical}%
  \BibitemOpen
  \bibfield  {author} {\bibinfo {author} {\bibfnamefont {A.}~\bibnamefont
  {Rai}}, \bibinfo {author} {\bibfnamefont {A.}~\bibnamefont {Mahata}},
  \bibinfo {author} {\bibfnamefont {M.}~\bibnamefont {Nurujjaman}}, \ and\
  \bibinfo {author} {\bibfnamefont {O.}~\bibnamefont {Prakash}},\ }\href@noop
  {} {\bibfield  {journal} {\bibinfo  {journal} {International Journal of
  Modern Physics C}\ }\textbf {\bibinfo {volume} {33}},\ \bibinfo {pages}
  {2250019} (\bibinfo {year} {2022}{\natexlab{a}})}\BibitemShut {NoStop}%
\bibitem [{\citenamefont {Pal}(2010)}]{pal2010foreign}%
  \BibitemOpen
  \bibfield  {author} {\bibinfo {author} {\bibfnamefont {P.}~\bibnamefont
  {Pal}},\ }\href@noop {} {\bibfield  {journal} {\bibinfo  {journal} {Capital
  Without Borders: Challenges to Development}\ }\textbf {\bibinfo {volume}
  {1}},\ \bibinfo {pages} {121} (\bibinfo {year} {2010})}\BibitemShut {NoStop}%
\bibitem [{\citenamefont {Nagaraj}(1996)}]{nagaraj1996india}%
  \BibitemOpen
  \bibfield  {author} {\bibinfo {author} {\bibfnamefont {R.}~\bibnamefont
  {Nagaraj}},\ }\href@noop {} {\bibfield  {journal} {\bibinfo  {journal}
  {Economic and Political Weekly}\ ,\ \bibinfo {pages} {2553}} (\bibinfo {year}
  {1996})}\BibitemShut {NoStop}%
\bibitem [{\citenamefont {Seyhun}(1990)}]{seyhun1990overreaction}%
  \BibitemOpen
  \bibfield  {author} {\bibinfo {author} {\bibfnamefont {H.~N.}\ \bibnamefont
  {Seyhun}},\ }\href@noop {} {\bibfield  {journal} {\bibinfo  {journal} {The
  Journal of Finance}\ }\textbf {\bibinfo {volume} {45}},\ \bibinfo {pages}
  {1363} (\bibinfo {year} {1990})}\BibitemShut {NoStop}%
\bibitem [{\citenamefont {Morris}\ and\ \citenamefont
  {Alam}(2012)}]{morris2012value}%
  \BibitemOpen
  \bibfield  {author} {\bibinfo {author} {\bibfnamefont {J.~J.}\ \bibnamefont
  {Morris}}\ and\ \bibinfo {author} {\bibfnamefont {P.}~\bibnamefont {Alam}},\
  }\href@noop {} {\bibfield  {journal} {\bibinfo  {journal} {The Quarterly
  Review of Economics and Finance}\ }\textbf {\bibinfo {volume} {52}},\
  \bibinfo {pages} {243} (\bibinfo {year} {2012})}\BibitemShut {NoStop}%
\bibitem [{\citenamefont {Acharya}\ and\ \citenamefont
  {Richardson}(2009)}]{acharya2009causes}%
  \BibitemOpen
  \bibfield  {author} {\bibinfo {author} {\bibfnamefont {V.~V.}\ \bibnamefont
  {Acharya}}\ and\ \bibinfo {author} {\bibfnamefont {M.}~\bibnamefont
  {Richardson}},\ }\href@noop {} {\bibfield  {journal} {\bibinfo  {journal}
  {Critical review}\ }\textbf {\bibinfo {volume} {21}},\ \bibinfo {pages} {195}
  (\bibinfo {year} {2009})}\BibitemShut {NoStop}%
\bibitem [{\citenamefont {Mahata}\ \emph
  {et~al.}(2021{\natexlab{b}})\citenamefont {Mahata}, \citenamefont {Rai},
  \citenamefont {Nurujjaman},\ and\ \citenamefont
  {Prakash}}]{mahata2021modeling}%
  \BibitemOpen
  \bibfield  {author} {\bibinfo {author} {\bibfnamefont {A.}~\bibnamefont
  {Mahata}}, \bibinfo {author} {\bibfnamefont {A.}~\bibnamefont {Rai}},
  \bibinfo {author} {\bibfnamefont {M.}~\bibnamefont {Nurujjaman}}, \ and\
  \bibinfo {author} {\bibfnamefont {O.}~\bibnamefont {Prakash}},\ }\href@noop
  {} {\bibfield  {journal} {\bibinfo  {journal} {Physica A: Statistical
  Mechanics and its Applications}\ }\textbf {\bibinfo {volume} {574}},\
  \bibinfo {pages} {126008} (\bibinfo {year} {2021}{\natexlab{b}})}\BibitemShut
  {NoStop}%
\bibitem [{\citenamefont {Rai}\ \emph {et~al.}(2022{\natexlab{b}})\citenamefont
  {Rai}, \citenamefont {Mahata}, \citenamefont {Nurujjaman}, \citenamefont
  {Majhi},\ and\ \citenamefont {Debnath}}]{rai2022sentiment}%
  \BibitemOpen
  \bibfield  {author} {\bibinfo {author} {\bibfnamefont {A.}~\bibnamefont
  {Rai}}, \bibinfo {author} {\bibfnamefont {A.}~\bibnamefont {Mahata}},
  \bibinfo {author} {\bibfnamefont {M.}~\bibnamefont {Nurujjaman}}, \bibinfo
  {author} {\bibfnamefont {S.}~\bibnamefont {Majhi}}, \ and\ \bibinfo {author}
  {\bibfnamefont {K.}~\bibnamefont {Debnath}},\ }\href@noop {} {\bibfield
  {journal} {\bibinfo  {journal} {Physica A: Statistical Mechanics and its
  Applications}\ }\textbf {\bibinfo {volume} {592}},\ \bibinfo {pages} {126810}
  (\bibinfo {year} {2022}{\natexlab{b}})}\BibitemShut {NoStop}%
\bibitem [{mon()}]{monsoon}%
  \BibitemOpen
  \href@noop {} {}\bibinfo {howpublished}
  {\url{https://blog.finology.in/stock-market/indian-stock-market-biggest-bull-runs}}\BibitemShut
  {NoStop}%
\bibitem [{\citenamefont {Gursida}(2017)}]{gursida2017influence}%
  \BibitemOpen
  \bibfield  {author} {\bibinfo {author} {\bibfnamefont {H.}~\bibnamefont
  {Gursida}},\ }\href@noop {} {\bibfield  {journal} {\bibinfo  {journal}
  {Jurnal Terapan Manajemen dan Bisnis}\ }\textbf {\bibinfo {volume} {3}},\
  \bibinfo {pages} {222} (\bibinfo {year} {2017})}\BibitemShut {NoStop}%
\bibitem [{\citenamefont {Emamgholipour}\ \emph {et~al.}(2013)\citenamefont
  {Emamgholipour}, \citenamefont {Pouraghajan}, \citenamefont {Tabari},
  \citenamefont {Haghparast},\ and\ \citenamefont
  {Shirsavar}}]{emamgholipour2013effects}%
  \BibitemOpen
  \bibfield  {author} {\bibinfo {author} {\bibfnamefont {M.}~\bibnamefont
  {Emamgholipour}}, \bibinfo {author} {\bibfnamefont {A.}~\bibnamefont
  {Pouraghajan}}, \bibinfo {author} {\bibfnamefont {N.~A.~Y.}\ \bibnamefont
  {Tabari}}, \bibinfo {author} {\bibfnamefont {M.}~\bibnamefont {Haghparast}},
  \ and\ \bibinfo {author} {\bibfnamefont {A.~A.~A.}\ \bibnamefont
  {Shirsavar}},\ }\href@noop {} {\bibfield  {journal} {\bibinfo  {journal}
  {International Research Journal of Applied and Basic Sciences}\ }\textbf
  {\bibinfo {volume} {4}},\ \bibinfo {pages} {696} (\bibinfo {year}
  {2013})}\BibitemShut {NoStop}%
\bibitem [{\citenamefont {Wang}\ \emph {et~al.}(2013)\citenamefont {Wang},
  \citenamefont {Fu},\ and\ \citenamefont {Luo}}]{wang2013accounting}%
  \BibitemOpen
  \bibfield  {author} {\bibinfo {author} {\bibfnamefont {J.}~\bibnamefont
  {Wang}}, \bibinfo {author} {\bibfnamefont {G.}~\bibnamefont {Fu}}, \ and\
  \bibinfo {author} {\bibfnamefont {C.}~\bibnamefont {Luo}},\ }\href@noop {}
  {\bibfield  {journal} {\bibinfo  {journal} {Journal of Business \&
  Management}\ }\textbf {\bibinfo {volume} {2}},\ \bibinfo {pages} {11}
  (\bibinfo {year} {2013})}\BibitemShut {NoStop}%
\bibitem [{\citenamefont {Cutler}\ \emph {et~al.}(1988)\citenamefont {Cutler},
  \citenamefont {Poterba},\ and\ \citenamefont {Summers}}]{cutler1988moves}%
  \BibitemOpen
  \bibfield  {author} {\bibinfo {author} {\bibfnamefont {D.~M.}\ \bibnamefont
  {Cutler}}, \bibinfo {author} {\bibfnamefont {J.~M.}\ \bibnamefont {Poterba}},
  \ and\ \bibinfo {author} {\bibfnamefont {L.~H.}\ \bibnamefont {Summers}},\
  }\href@noop {} {\enquote {\bibinfo {title} {What moves stock prices?}}\ }
  (\bibinfo {year} {1988})\BibitemShut {NoStop}%
\bibitem [{\citenamefont {Patell}\ and\ \citenamefont
  {Wolfson}(1982)}]{patell1982good}%
  \BibitemOpen
  \bibfield  {author} {\bibinfo {author} {\bibfnamefont {J.~M.}\ \bibnamefont
  {Patell}}\ and\ \bibinfo {author} {\bibfnamefont {M.~A.}\ \bibnamefont
  {Wolfson}},\ }\href@noop {} {\bibfield  {journal} {\bibinfo  {journal}
  {Accounting review}\ }\textbf {\bibinfo {volume} {57}},\ \bibinfo {pages}
  {509} (\bibinfo {year} {1982})}\BibitemShut {NoStop}%
\bibitem [{\citenamefont {Taylor}\ and\ \citenamefont
  {Allen}(1992)}]{taylor1992use}%
  \BibitemOpen
  \bibfield  {author} {\bibinfo {author} {\bibfnamefont {M.~P.}\ \bibnamefont
  {Taylor}}\ and\ \bibinfo {author} {\bibfnamefont {H.}~\bibnamefont {Allen}},\
  }\href@noop {} {\bibfield  {journal} {\bibinfo  {journal} {Journal of
  international Money and Finance}\ }\textbf {\bibinfo {volume} {11}},\
  \bibinfo {pages} {304} (\bibinfo {year} {1992})}\BibitemShut {NoStop}%
\bibitem [{\citenamefont {Wong}\ \emph {et~al.}(2003)\citenamefont {Wong},
  \citenamefont {Manzur},\ and\ \citenamefont {Chew}}]{wong2003rewarding}%
  \BibitemOpen
  \bibfield  {author} {\bibinfo {author} {\bibfnamefont {W.-K.}\ \bibnamefont
  {Wong}}, \bibinfo {author} {\bibfnamefont {M.}~\bibnamefont {Manzur}}, \ and\
  \bibinfo {author} {\bibfnamefont {B.-K.}\ \bibnamefont {Chew}},\ }\href@noop
  {} {\bibfield  {journal} {\bibinfo  {journal} {Applied Financial Economics}\
  }\textbf {\bibinfo {volume} {13}},\ \bibinfo {pages} {543} (\bibinfo {year}
  {2003})}\BibitemShut {NoStop}%
\bibitem [{\citenamefont {Barlevy}\ and\ \citenamefont
  {Veronesi}(2003)}]{barlevy2003rational}%
  \BibitemOpen
  \bibfield  {author} {\bibinfo {author} {\bibfnamefont {G.}~\bibnamefont
  {Barlevy}}\ and\ \bibinfo {author} {\bibfnamefont {P.}~\bibnamefont
  {Veronesi}},\ }\href@noop {} {\bibfield  {journal} {\bibinfo  {journal}
  {Journal of Economic Theory}\ }\textbf {\bibinfo {volume} {110}},\ \bibinfo
  {pages} {234} (\bibinfo {year} {2003})}\BibitemShut {NoStop}%
\bibitem [{\citenamefont {Gennotte}\ and\ \citenamefont
  {Leland}(1990)}]{gennotte1990market}%
  \BibitemOpen
  \bibfield  {author} {\bibinfo {author} {\bibfnamefont {G.}~\bibnamefont
  {Gennotte}}\ and\ \bibinfo {author} {\bibfnamefont {H.}~\bibnamefont
  {Leland}},\ }\href@noop {} {\bibfield  {journal} {\bibinfo  {journal} {The
  American Economic Review}\ ,\ \bibinfo {pages} {999}} (\bibinfo {year}
  {1990})}\BibitemShut {NoStop}%
\bibitem [{\citenamefont {Leigh}\ \emph {et~al.}(2002)\citenamefont {Leigh},
  \citenamefont {Modani}, \citenamefont {Purvis},\ and\ \citenamefont
  {Roberts}}]{leigh2002stock}%
  \BibitemOpen
  \bibfield  {author} {\bibinfo {author} {\bibfnamefont {W.}~\bibnamefont
  {Leigh}}, \bibinfo {author} {\bibfnamefont {N.}~\bibnamefont {Modani}},
  \bibinfo {author} {\bibfnamefont {R.}~\bibnamefont {Purvis}}, \ and\ \bibinfo
  {author} {\bibfnamefont {T.}~\bibnamefont {Roberts}},\ }\href@noop {}
  {\bibfield  {journal} {\bibinfo  {journal} {Expert Systems with
  Applications}\ }\textbf {\bibinfo {volume} {23}},\ \bibinfo {pages} {155}
  (\bibinfo {year} {2002})}\BibitemShut {NoStop}%
\bibitem [{\citenamefont {Bulkowski}(2021)}]{bulkowski2021encyclopedia}%
  \BibitemOpen
  \bibfield  {author} {\bibinfo {author} {\bibfnamefont {T.~N.}\ \bibnamefont
  {Bulkowski}},\ }\href@noop {} {\emph {\bibinfo {title} {Encyclopedia of chart
  patterns}}}\ (\bibinfo  {publisher} {John Wiley \& Sons},\ \bibinfo {year}
  {2021})\BibitemShut {NoStop}%
\bibitem [{\citenamefont {Petersen}\ \emph {et~al.}(2010)\citenamefont
  {Petersen}, \citenamefont {Wang}, \citenamefont {Havlin},\ and\ \citenamefont
  {Stanley}}]{petersen2010quantitative}%
  \BibitemOpen
  \bibfield  {author} {\bibinfo {author} {\bibfnamefont {A.~M.}\ \bibnamefont
  {Petersen}}, \bibinfo {author} {\bibfnamefont {F.}~\bibnamefont {Wang}},
  \bibinfo {author} {\bibfnamefont {S.}~\bibnamefont {Havlin}}, \ and\ \bibinfo
  {author} {\bibfnamefont {H.~E.}\ \bibnamefont {Stanley}},\ }\href@noop {}
  {\bibfield  {journal} {\bibinfo  {journal} {Physical Review E}\ }\textbf
  {\bibinfo {volume} {81}},\ \bibinfo {pages} {066121} (\bibinfo {year}
  {2010})}\BibitemShut {NoStop}%
\bibitem [{\citenamefont {Nazir}\ \emph {et~al.}(2014)\citenamefont {Nazir},
  \citenamefont {Younus}, \citenamefont {Kaleem},\ and\ \citenamefont
  {Anwar}}]{nazir2014impact}%
  \BibitemOpen
  \bibfield  {author} {\bibinfo {author} {\bibfnamefont {M.~S.}\ \bibnamefont
  {Nazir}}, \bibinfo {author} {\bibfnamefont {H.}~\bibnamefont {Younus}},
  \bibinfo {author} {\bibfnamefont {A.}~\bibnamefont {Kaleem}}, \ and\ \bibinfo
  {author} {\bibfnamefont {Z.}~\bibnamefont {Anwar}},\ }\href@noop {}
  {\bibfield  {journal} {\bibinfo  {journal} {Journal of Economic and
  Administrative Sciences}\ } (\bibinfo {year} {2014})}\BibitemShut {NoStop}%
\bibitem [{\citenamefont {Mahmood}\ \emph {et~al.}(2014)\citenamefont
  {Mahmood}, \citenamefont {Irfan}, \citenamefont {Iqbal}, \citenamefont
  {Kamran},\ and\ \citenamefont {Ijaz}}]{mahmood2014impact}%
  \BibitemOpen
  \bibfield  {author} {\bibinfo {author} {\bibfnamefont {S.}~\bibnamefont
  {Mahmood}}, \bibinfo {author} {\bibfnamefont {M.}~\bibnamefont {Irfan}},
  \bibinfo {author} {\bibfnamefont {S.}~\bibnamefont {Iqbal}}, \bibinfo
  {author} {\bibfnamefont {M.}~\bibnamefont {Kamran}}, \ and\ \bibinfo {author}
  {\bibfnamefont {A.}~\bibnamefont {Ijaz}},\ }\href@noop {} {\bibfield
  {journal} {\bibinfo  {journal} {Journal of Asian Business Strategy}\ }\textbf
  {\bibinfo {volume} {4}},\ \bibinfo {pages} {163} (\bibinfo {year}
  {2014})}\BibitemShut {NoStop}%
\bibitem [{\citenamefont {Boungou}\ and\ \citenamefont
  {Yati{\'e}}(2022)}]{boungou2022impact}%
  \BibitemOpen
  \bibfield  {author} {\bibinfo {author} {\bibfnamefont {W.}~\bibnamefont
  {Boungou}}\ and\ \bibinfo {author} {\bibfnamefont {A.}~\bibnamefont
  {Yati{\'e}}},\ }\href@noop {} {\bibfield  {journal} {\bibinfo  {journal}
  {Economics Letters}\ }\textbf {\bibinfo {volume} {215}},\ \bibinfo {pages}
  {110516} (\bibinfo {year} {2022})}\BibitemShut {NoStop}%
\bibitem [{\citenamefont {Hudson}\ and\ \citenamefont
  {Urquhart}(2015)}]{hudson2015war}%
  \BibitemOpen
  \bibfield  {author} {\bibinfo {author} {\bibfnamefont {R.}~\bibnamefont
  {Hudson}}\ and\ \bibinfo {author} {\bibfnamefont {A.}~\bibnamefont
  {Urquhart}},\ }\href@noop {} {\bibfield  {journal} {\bibinfo  {journal}
  {International Review of Financial Analysis}\ }\textbf {\bibinfo {volume}
  {40}},\ \bibinfo {pages} {166} (\bibinfo {year} {2015})}\BibitemShut
  {NoStop}%
\bibitem [{\citenamefont {Schneider}\ and\ \citenamefont
  {Troeger}(2006)}]{schneider2006war}%
  \BibitemOpen
  \bibfield  {author} {\bibinfo {author} {\bibfnamefont {G.}~\bibnamefont
  {Schneider}}\ and\ \bibinfo {author} {\bibfnamefont {V.~E.}\ \bibnamefont
  {Troeger}},\ }\href@noop {} {\bibfield  {journal} {\bibinfo  {journal}
  {Journal of conflict resolution}\ }\textbf {\bibinfo {volume} {50}},\
  \bibinfo {pages} {623} (\bibinfo {year} {2006})}\BibitemShut {NoStop}%
\bibitem [{\citenamefont {Worthington*}\ and\ \citenamefont
  {Valadkhani}(2004)}]{worthington2004measuring}%
  \BibitemOpen
  \bibfield  {author} {\bibinfo {author} {\bibfnamefont {A.}~\bibnamefont
  {Worthington*}}\ and\ \bibinfo {author} {\bibfnamefont {A.}~\bibnamefont
  {Valadkhani}},\ }\href@noop {} {\bibfield  {journal} {\bibinfo  {journal}
  {Applied Economics}\ }\textbf {\bibinfo {volume} {36}},\ \bibinfo {pages}
  {2177} (\bibinfo {year} {2004})}\BibitemShut {NoStop}%
\bibitem [{\citenamefont {Bourdeau-Brien}\ and\ \citenamefont
  {Kryzanowski}(2017)}]{bourdeau2017impact}%
  \BibitemOpen
  \bibfield  {author} {\bibinfo {author} {\bibfnamefont {M.}~\bibnamefont
  {Bourdeau-Brien}}\ and\ \bibinfo {author} {\bibfnamefont {L.}~\bibnamefont
  {Kryzanowski}},\ }\href@noop {} {\bibfield  {journal} {\bibinfo  {journal}
  {The Quarterly Review of Economics and Finance}\ }\textbf {\bibinfo {volume}
  {63}},\ \bibinfo {pages} {259} (\bibinfo {year} {2017})}\BibitemShut
  {NoStop}%
\bibitem [{\citenamefont {Mishra}\ \emph {et~al.}(2020)\citenamefont {Mishra},
  \citenamefont {Leo~Kingston}, \citenamefont {Hens}, \citenamefont
  {Kapitaniak}, \citenamefont {Feudel},\ and\ \citenamefont
  {Dana}}]{mishra2020routes}%
  \BibitemOpen
  \bibfield  {author} {\bibinfo {author} {\bibfnamefont {A.}~\bibnamefont
  {Mishra}}, \bibinfo {author} {\bibfnamefont {S.}~\bibnamefont
  {Leo~Kingston}}, \bibinfo {author} {\bibfnamefont {C.}~\bibnamefont {Hens}},
  \bibinfo {author} {\bibfnamefont {T.}~\bibnamefont {Kapitaniak}}, \bibinfo
  {author} {\bibfnamefont {U.}~\bibnamefont {Feudel}}, \ and\ \bibinfo {author}
  {\bibfnamefont {S.~K.}\ \bibnamefont {Dana}},\ }\href@noop {} {\bibfield
  {journal} {\bibinfo  {journal} {Chaos: An Interdisciplinary Journal of
  Nonlinear Science}\ }\textbf {\bibinfo {volume} {30}},\ \bibinfo {pages}
  {063114} (\bibinfo {year} {2020})}\BibitemShut {NoStop}%
\bibitem [{\citenamefont {Chaurasia}\ \emph {et~al.}(2020)\citenamefont
  {Chaurasia}, \citenamefont {Verma},\ and\ \citenamefont
  {Sinha}}]{chaurasia2020advent}%
  \BibitemOpen
  \bibfield  {author} {\bibinfo {author} {\bibfnamefont {S.~S.}\ \bibnamefont
  {Chaurasia}}, \bibinfo {author} {\bibfnamefont {U.~K.}\ \bibnamefont
  {Verma}}, \ and\ \bibinfo {author} {\bibfnamefont {S.}~\bibnamefont
  {Sinha}},\ }\href@noop {} {\bibfield  {journal} {\bibinfo  {journal}
  {Scientific Reports}\ }\textbf {\bibinfo {volume} {10}},\ \bibinfo {pages}
  {1} (\bibinfo {year} {2020})}\BibitemShut {NoStop}%
\bibitem [{\citenamefont {Kim}(2003)}]{kim2003financial}%
  \BibitemOpen
  \bibfield  {author} {\bibinfo {author} {\bibfnamefont {K.-j.}\ \bibnamefont
  {Kim}},\ }\href@noop {} {\bibfield  {journal} {\bibinfo  {journal}
  {Neurocomputing}\ }\textbf {\bibinfo {volume} {55}},\ \bibinfo {pages} {307}
  (\bibinfo {year} {2003})}\BibitemShut {NoStop}%
\bibitem [{\citenamefont {Khaidem}\ \emph {et~al.}(2016)\citenamefont
  {Khaidem}, \citenamefont {Saha},\ and\ \citenamefont
  {Dey}}]{khaidem2016predicting}%
  \BibitemOpen
  \bibfield  {author} {\bibinfo {author} {\bibfnamefont {L.}~\bibnamefont
  {Khaidem}}, \bibinfo {author} {\bibfnamefont {S.}~\bibnamefont {Saha}}, \
  and\ \bibinfo {author} {\bibfnamefont {S.~R.}\ \bibnamefont {Dey}},\
  }\href@noop {} {\bibfield  {journal} {\bibinfo  {journal} {arXiv preprint
  arXiv:1605.00003}\ } (\bibinfo {year} {2016})}\BibitemShut {NoStop}%
\bibitem [{\citenamefont {Huang}\ and\ \citenamefont
  {Tsai}(2009)}]{huang2009hybrid}%
  \BibitemOpen
  \bibfield  {author} {\bibinfo {author} {\bibfnamefont {C.-L.}\ \bibnamefont
  {Huang}}\ and\ \bibinfo {author} {\bibfnamefont {C.-Y.}\ \bibnamefont
  {Tsai}},\ }\href@noop {} {\bibfield  {journal} {\bibinfo  {journal} {Expert
  Systems with applications}\ }\textbf {\bibinfo {volume} {36}},\ \bibinfo
  {pages} {1529} (\bibinfo {year} {2009})}\BibitemShut {NoStop}%
\bibitem [{\citenamefont {Kumbure}\ \emph {et~al.}(2022)\citenamefont
  {Kumbure}, \citenamefont {Lohrmann}, \citenamefont {Luukka},\ and\
  \citenamefont {Porras}}]{kumbure2022machine}%
  \BibitemOpen
  \bibfield  {author} {\bibinfo {author} {\bibfnamefont {M.~M.}\ \bibnamefont
  {Kumbure}}, \bibinfo {author} {\bibfnamefont {C.}~\bibnamefont {Lohrmann}},
  \bibinfo {author} {\bibfnamefont {P.}~\bibnamefont {Luukka}}, \ and\ \bibinfo
  {author} {\bibfnamefont {J.}~\bibnamefont {Porras}},\ }\href@noop {}
  {\bibfield  {journal} {\bibinfo  {journal} {Expert Systems with
  Applications}\ ,\ \bibinfo {pages} {116659}} (\bibinfo {year}
  {2022})}\BibitemShut {NoStop}%
\bibitem [{\citenamefont {Mushtaq}(2011)}]{mushtaq2011augmented}%
  \BibitemOpen
  \bibfield  {author} {\bibinfo {author} {\bibfnamefont {R.}~\bibnamefont
  {Mushtaq}},\ }\href@noop {} {\  (\bibinfo {year} {2011})}\BibitemShut
  {NoStop}%
\bibitem [{\citenamefont {Said}\ and\ \citenamefont
  {Dickey}(1984)}]{said1984testing}%
  \BibitemOpen
  \bibfield  {author} {\bibinfo {author} {\bibfnamefont {S.~E.}\ \bibnamefont
  {Said}}\ and\ \bibinfo {author} {\bibfnamefont {D.~A.}\ \bibnamefont
  {Dickey}},\ }\href@noop {} {\bibfield  {journal} {\bibinfo  {journal}
  {Biometrika}\ }\textbf {\bibinfo {volume} {71}},\ \bibinfo {pages} {599}
  (\bibinfo {year} {1984})}\BibitemShut {NoStop}%
\bibitem [{\citenamefont {Herranz}(2017)}]{herranz2017unit}%
  \BibitemOpen
  \bibfield  {author} {\bibinfo {author} {\bibfnamefont {E.}~\bibnamefont
  {Herranz}},\ }\href@noop {} {\bibfield  {journal} {\bibinfo  {journal} {Wiley
  Interdisciplinary Reviews: Computational Statistics}\ }\textbf {\bibinfo
  {volume} {9}},\ \bibinfo {pages} {e1396} (\bibinfo {year}
  {2017})}\BibitemShut {NoStop}%
\bibitem [{\citenamefont {Paparoditis}\ and\ \citenamefont
  {Politis}(2018)}]{paparoditis2018asymptotic}%
  \BibitemOpen
  \bibfield  {author} {\bibinfo {author} {\bibfnamefont {E.}~\bibnamefont
  {Paparoditis}}\ and\ \bibinfo {author} {\bibfnamefont {D.~N.}\ \bibnamefont
  {Politis}},\ }\href@noop {} {\bibfield  {journal} {\bibinfo  {journal}
  {Econometric Reviews}\ }\textbf {\bibinfo {volume} {37}},\ \bibinfo {pages}
  {955} (\bibinfo {year} {2018})}\BibitemShut {NoStop}%
\bibitem [{\citenamefont {Kim}\ and\ \citenamefont {Choi}(2017)}]{kim2017unit}%
  \BibitemOpen
  \bibfield  {author} {\bibinfo {author} {\bibfnamefont {J.~H.}\ \bibnamefont
  {Kim}}\ and\ \bibinfo {author} {\bibfnamefont {I.}~\bibnamefont {Choi}},\
  }\href@noop {} {\bibfield  {journal} {\bibinfo  {journal} {Econometrics}\
  }\textbf {\bibinfo {volume} {5}},\ \bibinfo {pages} {41} (\bibinfo {year}
  {2017})}\BibitemShut {NoStop}%
\bibitem [{\citenamefont {Huang}\ \emph {et~al.}(1998)\citenamefont {Huang},
  \citenamefont {Shen}, \citenamefont {Long}, \citenamefont {Wu}, \citenamefont
  {Shih}, \citenamefont {Zheng}, \citenamefont {Yen}, \citenamefont {Tung},\
  and\ \citenamefont {Liu}}]{huang1998empirical}%
  \BibitemOpen
  \bibfield  {author} {\bibinfo {author} {\bibfnamefont {N.~E.}\ \bibnamefont
  {Huang}}, \bibinfo {author} {\bibfnamefont {Z.}~\bibnamefont {Shen}},
  \bibinfo {author} {\bibfnamefont {S.~R.}\ \bibnamefont {Long}}, \bibinfo
  {author} {\bibfnamefont {M.~C.}\ \bibnamefont {Wu}}, \bibinfo {author}
  {\bibfnamefont {H.~H.}\ \bibnamefont {Shih}}, \bibinfo {author}
  {\bibfnamefont {Q.}~\bibnamefont {Zheng}}, \bibinfo {author} {\bibfnamefont
  {N.-C.}\ \bibnamefont {Yen}}, \bibinfo {author} {\bibfnamefont {C.~C.}\
  \bibnamefont {Tung}}, \ and\ \bibinfo {author} {\bibfnamefont {H.~H.}\
  \bibnamefont {Liu}},\ }\href@noop {} {\bibfield  {journal} {\bibinfo
  {journal} {Proceedings of the Royal Society of London. Series A:
  mathematical, physical and engineering sciences}\ }\textbf {\bibinfo {volume}
  {454}},\ \bibinfo {pages} {903} (\bibinfo {year} {1998})}\BibitemShut
  {NoStop}%
\bibitem [{\citenamefont {Mahata}\ \emph {et~al.}(2020)\citenamefont {Mahata},
  \citenamefont {Bal},\ and\ \citenamefont
  {Nurujjaman}}]{mahata2020identification}%
  \BibitemOpen
  \bibfield  {author} {\bibinfo {author} {\bibfnamefont {A.}~\bibnamefont
  {Mahata}}, \bibinfo {author} {\bibfnamefont {D.~P.}\ \bibnamefont {Bal}}, \
  and\ \bibinfo {author} {\bibfnamefont {M.}~\bibnamefont {Nurujjaman}},\
  }\href@noop {} {\bibfield  {journal} {\bibinfo  {journal} {Physica A:
  Statistical Mechanics and its Applications}\ }\textbf {\bibinfo {volume}
  {545}},\ \bibinfo {pages} {123612} (\bibinfo {year} {2020})}\BibitemShut
  {NoStop}%
\bibitem [{\citenamefont {Mahata}\ and\ \citenamefont
  {Nurujjaman}(2020)}]{mahata2020time}%
  \BibitemOpen
  \bibfield  {author} {\bibinfo {author} {\bibfnamefont {A.}~\bibnamefont
  {Mahata}}\ and\ \bibinfo {author} {\bibfnamefont {M.}~\bibnamefont
  {Nurujjaman}},\ }\href@noop {} {\bibfield  {journal} {\bibinfo  {journal}
  {Frontiers in Physics}\ }\textbf {\bibinfo {volume} {8}},\ \bibinfo {pages}
  {498} (\bibinfo {year} {2020})}\BibitemShut {NoStop}%
\bibitem [{\citenamefont {Karimipour}\ \emph {et~al.}(2019)\citenamefont
  {Karimipour}, \citenamefont {Bagherzadeh}, \citenamefont {Taghipour},
  \citenamefont {Abdollahi},\ and\ \citenamefont
  {Safaei}}]{karimipour2019novel}%
  \BibitemOpen
  \bibfield  {author} {\bibinfo {author} {\bibfnamefont {A.}~\bibnamefont
  {Karimipour}}, \bibinfo {author} {\bibfnamefont {S.~A.}\ \bibnamefont
  {Bagherzadeh}}, \bibinfo {author} {\bibfnamefont {A.}~\bibnamefont
  {Taghipour}}, \bibinfo {author} {\bibfnamefont {A.}~\bibnamefont
  {Abdollahi}}, \ and\ \bibinfo {author} {\bibfnamefont {M.~R.}\ \bibnamefont
  {Safaei}},\ }\href@noop {} {\bibfield  {journal} {\bibinfo  {journal}
  {Physica A: Statistical Mechanics and its Applications}\ }\textbf {\bibinfo
  {volume} {521}},\ \bibinfo {pages} {89} (\bibinfo {year} {2019})}\BibitemShut
  {NoStop}%
\bibitem [{Yah()}]{Yahoo}%
  \BibitemOpen
  \href@noop {} {}\bibinfo {howpublished}
  {\url{https://finance.yahoo.com/}}\BibitemShut {NoStop}%
\bibitem [{ish()}]{ishan}%
  \BibitemOpen
  \href@noop {} {}\bibinfo {howpublished}
  {\url{https://www.bseindia.com/xml-data/corpfiling/AttachHis/42e9b059-c135-4572-a02b-cbc7e511eced.pdf}}\BibitemShut
  {NoStop}%
\bibitem [{Rus()}]{Rusukraine}%
  \BibitemOpen
  \href@noop {} {}\bibinfo {howpublished}
  {\url{https://edition.cnn.com/europe/live-news/ukraine-russia-news-02-24-22-intl/index.html}}\BibitemShut
  {NoStop}%
\bibitem [{war()}]{war}%
  \BibitemOpen
  \href@noop {} {}\bibinfo {howpublished}
  {\url{https://economictimes.indiatimes.com/markets/stocks/news/russian-stock-market-crushed-by-war/-will-partially-reopen/articleshow/90413163.cms}}\BibitemShut
  {NoStop}%
\bibitem [{pnb()}]{pnb}%
  \BibitemOpen
  \href@noop {} {}\bibinfo {howpublished}
  {\url{https://www.business-standard.com/article/companies/carlyle-group-to-acquire-controlling-stake/-in-pnb-housing-finance-121060100058_1.html}}\BibitemShut
  {NoStop}%
\bibitem [{che()}]{chevron}%
  \BibitemOpen
  \href@noop {} {}\bibinfo {howpublished}
  {\url{https://www.entrepreneur.com/article/420500}}\BibitemShut {NoStop}%
\bibitem [{gam()}]{gamestop}%
  \BibitemOpen
  \href@noop {} {}\bibinfo {howpublished}
  {\url{https://www.nytimes.com/2021/01/27/business/gamestop-wall-street-bets.html}}\BibitemShut
  {NoStop}%
\bibitem [{net()}]{netflx}%
  \BibitemOpen
  \href@noop {} {}\bibinfo {howpublished}
  {\url{https://edition.cnn.com/2022/04/19/media/netflix-earnings/index.html}}\BibitemShut
  {NoStop}%
\bibitem [{pin()}]{pinduo}%
  \BibitemOpen
  \href@noop {} {}\bibinfo {howpublished}
  {\url{https://www.fool.com/investing/2022/03/16/why-baidu-tencent-holdings-and-pinduoduo-skyrocket/}}\BibitemShut
  {NoStop}%
\end{thebibliography}%

\end{document}